\documentclass[10pt]{article}
\usepackage{epsfig,amssymb,latexsym,color,amsmath,pifont}
\newcommand{\n}{\noindent} 
\newcommand{\e}{\varepsilon}
\newcommand{\R}{\mathbb{R}}
\newcommand{\inv}{invariant }
\newcommand{\x}{\mathbf{x}}
\newcommand{\T}{\mathbf{T}}
\newcommand{\Z}{\mathbb{Z}}
\newcommand{\fou}{\text{fou}}

\begin{document}
\twocolumn[

\title{\textcolor{blue}{\textbf{Ergodic Theory and Visualization I: Mesochronic Plots for Visualization of Ergodic Partition and Invariant Sets}}}
\author{Zoran Levnaji\'c${}^{1}$, Igor Mezi\'c${}^{2}$} 
\date{}
\maketitle

\begin{center}  \begin{minipage}{6.35in}

\begin{flushleft}
 ${}^{1}$\textit{Department of Physics and Astronomy, University of Potsdam, Karl-Liebknecht-Street 24/25, D-14476 Potsdam-Golm, Germany} \\
 ${}^{2}$\textit{Department of Mechanical Engineering, University of California Santa Barbara, Santa Barbara, CA 93106, USA} 
\end{flushleft}

We present a computational study of a  visualization method for invariant sets based on ergodic partition theory, first proposed in \cite{mezic:1994,mezicwiggy}. The algorithms for computation of the time averages of observables on phase space are developed and used to provide  an approximation of the ergodic partition of the phase space. We term the graphical representation of this approximation - based on time averages of observables - a Mesochronic Plot\footnote{From Greek: \textit{meso} - mean, \textit{chrono} - time}. The method is useful for identifying low-dimensional projections (e.g. two-dimensional slices) of invariant structures in phase spaces of dimensionality bigger than two. We also introduce the concept of the ergodic quotient space, obtained by assigning a point to every ergodic set, and provide an embedding method whose graphical representation we call the Mesochronic Scatter Plot (MSP). We use the Chirikov standard map as a well-known and dynamically rich example in order to illustrate the implementation of our methods. In addition, we expose applications to other higher dimensional maps such as the Fro\'eschle map for which we utilize our methods to analyze merging of resonances and, the three-dimensional Extended standard map for which we study the conjecture on its ergodicity \cite{esm}. We extend the study in our next paper \cite{levnajicmezic2} by investigating the visualization of periodic sets using harmonic time averages. Both of these methods are related to eigenspace structure of the Koopman operator. 
\end{minipage}  \end{center}

]


\n \textbf{Dynamical equations describing  behavior of systems of scientific interest are often impossible to solve analytically, and one must resort to various computational methods approximating the actual solution. We develop  computational methods for studying invariant sets of dynamical systems with arbitrary dimensionality using methods of ergodic partition theory. Our method consists in computing time averages of chosen functions under the phase space dynamics for a given trajectory. This allows us to obtain graphical visualization of \inv sets, including easily obtainable intersections of such sets with lower-dimensional surfaces or manifolds. The numerical efficiency of the method - interestingly - improves when the dynamics is more complex. To represent the obtained information in a condensed form, we introduce scatter plots that represent the global topology of ergodic sets. We illustrate the utility and extent of our methods by investigating the global dynamical properties of several known measure-preserving dynamical systems. In the context of dissipative systems, the method visualizes basins of attraction.}
\begin{center}
 ------------------------------------------------
\end{center}

\section{Introduction}

The increasing range and dimensionality of phenomena modeled as dynamical systems  creates the demand for new and better computational approaches to comprehend high-dimensional, complex dynamics. Most of the dynamical systems problems of  current scientific interest cannot be treated by purely analytical techniques, constraining one to chose among diverse computational methods \cite{Jones:2001}. The choice of the numerical method is primarily determined by the nature of the problem, but also by the type of investigation that is to be conducted -- the solution of the entire problem is often not required, and the attention is focused on the relevant results only. 

The choice of the computational approach can be the decisive factor for the overall efficiency of investigation and the precision of the results. A variety of methods are used for studying dynamical systems -- the direct numerical integration of equations of motion still remains the common approach, both for discrete-time (maps) and continuous-time (ODEs) systems. In this context, the visualization methods play a major role: it is often of interest to graphically visualize various aspects of motion, for instance the phase space structure of a given dynamics, without necessarily solving the entire system of equations. However, there is lack of such methods that easily extend to high-dimensional, complex nature of motion exhibited by many dynamical systems of interest. It would be useful, in particular, to have a method that enables visualization of low-dimensional (e.g. 2D) slices through higher dimensional invariant structures. We develop such a method, suggested first in \cite{mezic:1994,mezicwiggy}.

Visualization methods generally work as algorithms for dividing the dynamics' phase space into subsets according to a prescribed property of interest. \textit{Exit time plots} \cite{meiss} are computed by fixing a bounded subset of the phase space and measuring the time needed for its representative points to exit the subset, which are colored according to their exit-times. The phase space is then sliced in the equal-exit-time regions, giving conclusions regarding the system's transport properties. The \textit{set-oriented computational algorithms} \cite{dellnitz} allow visualization of the \inv sets and attractors by the appropriate phase space subdivision. After covering a tentative \inv region with boxes of certain (small) size, a sequence of box-size reductions is applied, and the optimal approximation of the \inv geometry is obtained at the limit. Similarly, by studying the \textit{return time dynamics} \cite{thiere} one constructs a phase space decomposition into almost \inv sets by measuring and comparing the return times for each subdivision element.  Methods of computing (un)stable manifolds based on analysis of \textit{geodesic level sets} were proposed in the context of vector fields  \cite{krauskopf}, having a particular importance for control problems. Invariant manifolds can also be visualized using distribution of finite-time Lyapunov exponents in phase space \cite{guzzo}, or by integration of \textit{fat trajectories} which extend the concept of standard numerical trajectory integration \cite{henderson-fat}.

On the other hand, ergodic theory \cite{walters} as a study of statistical aspects of motion is extensively used in the context of chaotic dynamical systems \cite{eckmannruelle}. Its applications  gave results that range from optimal mixing \cite{dalessandro}, traffic jams \cite{trafficergodic} and quantum many-body problems \cite{manybodyergodic}, to mathematical study of maximizing measures in discrete dynamics \cite{jenkinson}. Finally, study of properties of time averages and harmonic averages of dynamical systems gave rise to new  methods for visualization of invariant and periodic structures, both theoretically \cite{mezicwiggy,mezicbanaszuk} and experimentally \cite{mezicsoti}.

This paper presents a detailed computational study of the visualization method based on ergodic theory concepts proposed by Mezi\'c  in \cite{mezic:1994}. The method consists of computation of joint level sets of time averages of a basis of functions. This leads to an approximation to the \textit{ergodic partition} of the phase space. We name the resulting plot of joint level sets the Mesochronic Plot (since the method is based on time averaging). This method enables study of high-dimensional dynamical systems by visualizing the \inv set decomposition of appropriately chosen lower dimensional planes/surfaces that slice/intersect the examined phase space region. Moreover, the method can be similarly employed to visualize attractor basins for systems of arbitrary dimensionality. To  illustrate the implementation of our method we start with the Chirikov standard map \cite{chirikov} as a well-known example of a chaotic dynamical systems that possesses a rich variety of dynamical behaviors. We stress that the employment of Mesochronic Plots for 2D standard map is done for illustrative purposes only, while the real application of the method lies in systems with higher dimensionality such as the extended standard map and the Fro\'eschle map that we also study in this paper. 

We also introduce here the concept of the {\it ergodic quotient space}, where each ergodic set is associated with a point in the quotient space. We show how to approximately embed the resulting set into Euclidean space using time averages of observables. This leads to a novel type of a plot, called the Mesochronic Scatter Plot (MSP). We relate properties of MSP to dynamics in phase space.

As opposed to the return time statistics (yielding almost \inv sets) \cite{thiere} and exit time \cite{meiss} approaches which are dealing with the properties of transport and its speed, our method focuses on geometrical phase space properties. We are complementing these methods by constructing an algorithm that gives decomposition into ergodic sets (at the limit). In contrast to techniques involving finite-time Lyapunov exponents (similarly suitable for systems of arbitrary dimensionality) \cite{guzzo}, our method does not seek to identify finite-time structure of  normally hyperbolic invariant manifolds - which over time are typically densely embedded within the invariant structures that we define.

In our forthcoming paper \cite{levnajicmezic2} which relies on theoretical framework of Mezi\'c and Banaszuk \cite{mezicbanaszuk}, we employ harmonic time averages and develop an algorithm for visualization of periodic sets and resonances in the phase space. We show that our technique is related to known frequency map techniques which focus on the frequency spectrum of chosen trajectories developed by Laskar and collaborators \cite{laskar}, and relate such techniques to spectral properties (in particular, eigenfunctions) of the associated Koopman operator \cite{mezicbanaszuk}. However, while Laskar et al. plot the frequencies of motion, we plot the phases - thus unveiling the periodic (as opposed to invariant) partitions in the phase space. In addition, Laskar et al. methods are valid for near-integrable systems, and not rigorously supported for fully non-integrable systems. In contrast, our method have a rigorous justification for arbitrary measure-preserving systems. In relation to current work, while visualizing the periodic sets that resonate with a particular chosen frequency is of great importance, the technique of ergodic partitioning presented here provides a method of refining frequency partitioning of the entire phase space into even smaller invariant sets. Thus, frequency partitioning can in fact be thought as an example of a current, more general method.

This paper is organized as follows: after briefly discussing our method's theoretical background in Section \ref{The visualization method}, we show the construction of Mesochronic Plots for single functions under the standard map in Section \ref{Single-function plots}. The convergence of time averages is addressed in Section \ref{The Convergence Properties}. In Section \ref{The Clustering Methods and Visualization} we present the ergodic quotient space concept and its Mesochronic Scatter Plot embedding with multiple time averages, and construct a simple algorithm for ergodic partition approximation. In Sections \ref{The Froeschle Map} and \ref{Extended Standard Map} we expose concrete applications of our method to higher dimensional maps, like 4D Froeschl\'e map \cite{froeschle} and 3D extended standard map \cite{esm}. Conclusions are given in Section \ref{Conclusions}.


\section{The Visualization Method} \label{The visualization method}

In this Section we briefly sketch the mathematical basis of our visualization method and describe its implementation algorithm. As our approach here is rather application-oriented, we will skip the rigorous proofs and refer the reader to \cite{mezic:1994} (and references therein) for more details.

\subsection{The ergodic theory background} 

We consider a map $\T$ on a compact metric phase space $A$ endowed with a measure $\mu$ that is preserved under $\T$, evolving in discrete time $t$:  
\begin{equation}
 \x_{t+1} = \T \x_t, \;  t \in \Z, \ \ \x_t \in A. \label{invariance}
\end{equation}
Our central aim is to visualize \textit{\inv sets} $B \subset A$ for the map $\T$ defined as \cite{wiggy}:
\begin{equation}
\x_0  \in  B \;\; \Rightarrow \;\; \T^t \x_0 \in B \; \;\; \forall t \in \Z , \label{inv-set}
\end{equation}
which essentially means that each trajectory that starts in an \inv set $B$ stays in $B$ forever. There are many types of invariant sets
that play substantial role in analysis of dynamical systems. Here we are interested in {\it ergodic invariant sets}.

Consider $L^1$ real-valued functions on $A$ ($f:A\rightarrow \mathbb{R}\in L^1(A) \; \mbox{iff} \; \int_{A} |f (\mathbf{x})| d \mathbf{x} < \infty$), and let the \textit{time average} $f^*(\x_0)$ of a function $f \in L^1(A)$ corresponding to a phase space point $\x_0 \in A$ be defined as:
\begin{equation} 
f^* ({\bf x}_0) = \lim_{t \rightarrow \infty} \; \frac{1}{t}  \sum_{k=0}^{t-1}  f ({\bf T}^k {\bf x}_0) .  \label{ta-definition}
\end{equation}

By the Ergodic Theorem (\cite{walters}) this limit exists almost everywhere (a.e.) in $A$ for every $f \in L^1(A)$. We call the map $\T$ \textit{ergodic} over some set $B \subset A$ if for a.e. point $\x_0 \in B$ we have that the time average for every function $f \in L^1(B)$ is equal to the space average of that function $f$ over the set $B$. More precisely, ergodicity of $\T$ when restricted to $B$ (or ergodicity  over $B)$ means there exists an \textit{ergodic measure} $\mu_{B}$ such that the  equality: 
\begin{equation} f^* ({\bf x}_0) = \frac{1}{\mu_{B} (B)}  \int_{B} f d \mu_{B} , \end{equation}
holds a.e. in $B \subset A$. This also implies that $f^*$ is \textit{constant} a.e. in $B \subset A$ and that almost every orbit starting in $B$ covers $B$ densely at the limit; the amount of time a trajectory spends in a given region of $B$ is proportional to the ergodic measure of that region. Also, each $f^*$ is an \textit{\inv function} $f^* (\x_0) = f^*(\T^t \x_0) \; \forall n$. 
The set of real numbers $\R$ induces a partition of $A$ called $\zeta_f \equiv \{ B_a \}_{a \in \R}$ through a function $f \in L^1(A)$ :
\[ B_a= (f^*)^{-1}(a), \ \ \  \forall a \in \R   \]
where $(f^*)^{-1}$ is the set inverse of $f^*$, and some of the $B_a$ may be empty sets. We have: $$ \mu({\bigcup_a  B_a})=\mu(A), \ B_a\cap B_{a'}=\emptyset \; \forall a\neq a',$$ and, denoting the set of all the points for which the time average of $f$ does not exist by $\Sigma(f,\T)$,
\[  A=( {\cup_a  B_a})\bigcup \Sigma(f,\T). \] 
While this implies each $B_a$ to be \inv by construction, we may still have some set $B_a$ to actually be a union of more independent \inv sets that accidentally have the same time average for a given $f$. For this purpose we refine the partitioning by considering products of partitions corresponding to different  functions in order to obtain a partition in which each element is a non-decomposable, ergodic and \inv set. This final \textit{ergodic partition} $\zeta_e$ is defined as the product of all $\zeta_f$ belonging to a set ${\mathcal S} \subset L^1 (A)$: 
\begin{equation}
 \zeta_e = \bigvee_{f \in {\mathcal S}} \zeta_f ,  \label{ergodicpartition}
\end{equation}
where it suffices to take ${\mathcal S}$ to be a basis for $L^1 (A)$ \cite{mezicbanaszuk}.  Defining the Koopman operator $U$ associated with the map $\T$ by the composition operation
\[ Uf(\x)=f\circ \T(\x) \]
it is easy to see that,  the time average $f^*$ of a function $f$ is an eigenfunction of $U$ at eigenvalue $1$ \cite{mezicbanaszuk}.

The ergodic partition has the desired properties of dividing the phase space into a family of non-decomposable \inv sets: ergodic subsets are the ``minimal observable" \inv sets. In order to visualize the ergodic sets we need to approximate the ergodic partition by computing the time averages of a finite number of functions, and observe the subsets where they are simultaneously constant (i.e. the joint level sets of these time averages). In the rest of this work we will be developing and employing a computational algorithm that uses the described idea.

\subsection{The computational algorithm and numerical details}

Here we describe the algorithm for approximation of the ergodic partition, limiting for simplicity the discussion to the case of a 2D map with the phase space $A = [0,1] \times [0,1] \subset \R^2$.
\begin{description}
    \item[step 1] Set up a grid (e.g. lattice) of initial grid-points $(x_0,y_0)$ on the phase space $A$ 
    \item[step 2] Pick $N$ functions $\{ f_1, \hdots f_N \}$ from $L^1 (A)$ and compute their partial time averages for $t_{\mbox{final}}$
 iterations for each grid-point, which serve as the approximations for the real time averages $\{f^*_1, \hdots  f^*_N \}$
    \item[step 3] To every initial grid-point $(x_0,y_0)$ associate the corresponding \textit{time average vector}: 
\begin{equation} \begin{array}{ll}
& (x_0,y_0)  \longrightarrow  \mathbf{\bar f} (x_0,y_0), \\
& \mathbf{\bar f} (x_0,y_0) = \{f^*_1 (x_0,y_0),\hdots  f^*_N (x_0,y_0) \} \in \R^N
\end{array} \label{tavectors}
\end{equation}
    \item[step 4] Observe the distribution of time average vectors $\mathbf{\bar f}$ in $\R^N$ and group them optimally into clusters. Divide $A$ into a union of
 subsets, with each subset being given by those grid-points $(x_0,y_0)$ whose time average vectors $\mathbf{\bar f} (x_0,y_0)$ belong to the same cluster in $\R^N$. This family 
 of subsets is an $N$-function approximation of the ergodic partition of $A$ in the sense of Eq.\,(\ref{ergodicpartition}). 
\end{description}
The optimal number of iterations $t_{\mbox{final}}$ is to be set in accordance with the convergence properties of the time averages; observe also that the quality/properties of visualization strongly depend on the way time average vector are clustered: these topic will be dealt with later.

The choice of functions $\{ f_1, \hdots f_N \}$ is important: linearly dependent functions will not give  new information as their time averages differ only by a multiplicative constant. It is therefore necessary to consider linearly independent functions, which we shall do by picking them from an orthogonal basis on $L^1(A)$. 

We use a grid of $D \times D$ initial grid-points and iterate the dynamics for $t_{\mbox{final}}$ iterations, with values of $D$ and $t_{\mbox{final}}$ depending on the particular map/case under investigation.

\subsection{Standard map as the testing prototype}

We chose the Chirikov standard map \cite{chirikov} for testing the performance of our method. Its behavior has been widely studied and well-understood \cite{levnajictadic,venegeroles}. We consider it in the form: 
\begin{equation}
\begin{array}{lllc}
  x' &=& x + y  + \e \sin (2 \pi x)  \;\;\;  &[mod \; 1]  \\
  y' &=& y + \e \sin (2 \pi x)  \;\;\;  &[mod \; 1]
\end{array}
\label{mojasm}
\end{equation}
where $(x,y) \in [0,1] \times [0,1] \; \equiv [0,1]^2$ (the usual standard map's parameter $k$ is here $k=2\pi \e$).  It is an area-preserving (symplectic) map which exhibits a variety of \inv sets, both regular, composed of  periodic or  quasi-periodic orbits, and chaotic zones that evolve in size and structure as the parameter $\e$ is varied.


\section{Single-function Plots} \label{Single-function plots}

We set $N=1$ and consider the time averages of a single function under the dynamics of Eq.(\ref{mojasm}) for a grid of initial conditions. As the range of final time average values varies, we adjust the coloring scheme for each plot by assigning blue to the minimum and red to the maximum value obtained.

\subsection{The Fourier orthogonal basis}

\begin{figure*}[!hbt] \begin{center}
    \includegraphics[height=14.cm,width=12.5cm]{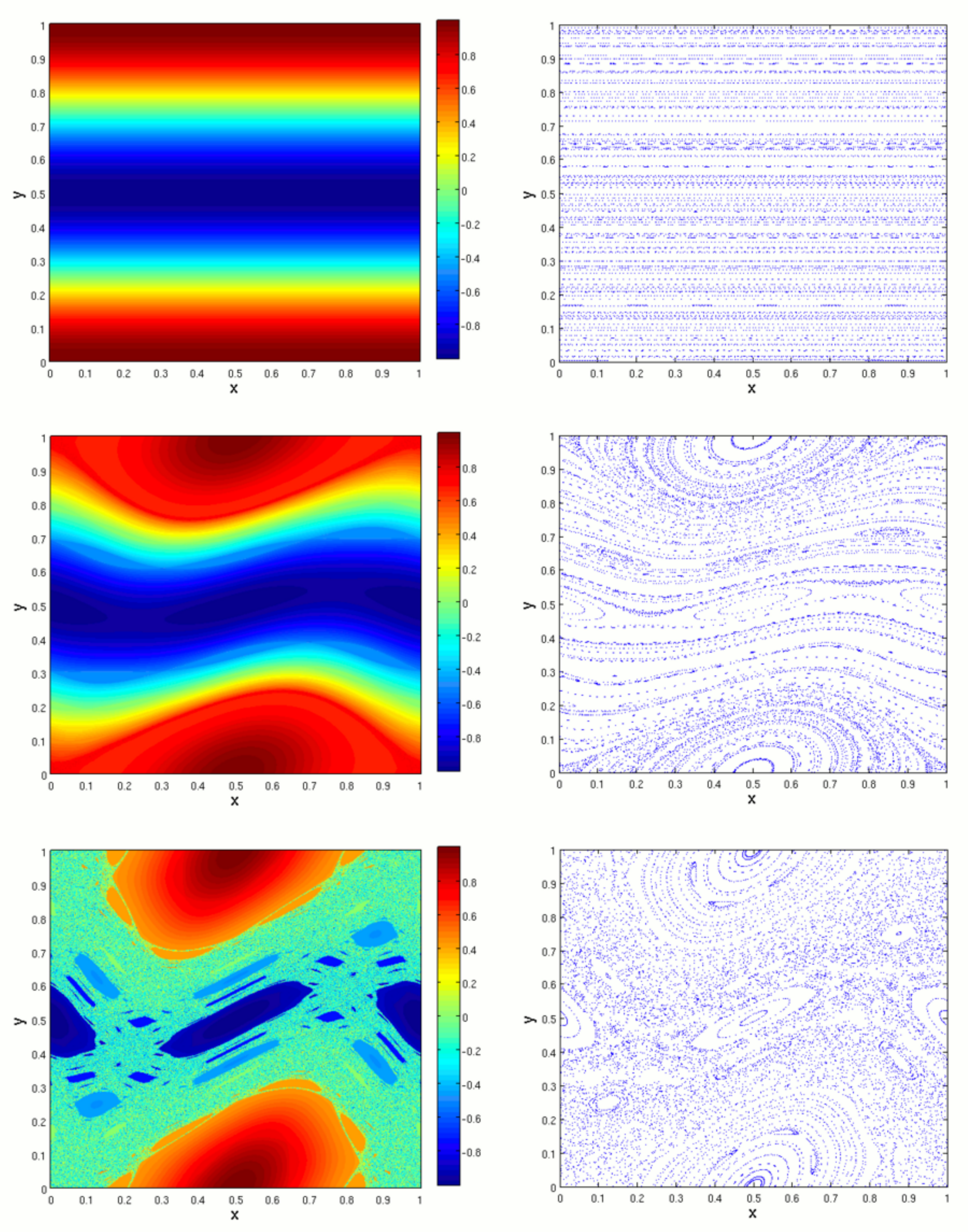}
    \caption{Single-function color-plots of time averages for the standard map Eq.(\ref{mojasm}) (left), and the corresponding phase space portraits (right) done as 100 iterations of 
  $11\times 11$ random trajectories picked from $[0,1]^2$, for the function $f=\cos(2\pi y)$. Top row: $\e=0$, middle:  $\e=0.09$, bottom: $\e=0.18$. The grid of $800 \times 800$ 
  initial points was used, and the dynamics was run for  $t_{\mbox{final}}=30000$ iterations.} \label{fourier} 
\end{center}  \end{figure*}

We pick the functions to time average  from the Fourier basis in the form:
\[   \fou_1 (2n\pi x) \; \fou_2(2m\pi y), \;\; \mbox{with} \;\; n,m \in \mathbb{N} , \label{fourierbasis}  \]
where $\fou_i = \sin$ or $\cos$, for $i=1,2$, obtaining a grid of time averages corresponding to the grid of initial conditions. The time average grid is then colored, visualizing the \inv sets as the uniformly colored level sets. In Fig.\,\ref{fourier} we show three examples of time average plots for various $\e$-values (left), along with their phase space portraits (right). The pictures on the top deal with $\e=0$ case, where the map possesses a family of \inv circles with $y=const.$ (a full measure of these circles have ergodic dynamics - irrational rotation - on them). Accordingly, the plot is a family of horizontal single-color lines, with color depending on the time average of the function over each $y=const.$ line. Middle figures show the case of $\e=0.09$ where the phase space is a mix between a small chaotic zone and families of regular islands. The bottom two figures regard the case of $\e=0.18$ characterized by the presence of a large chaotic zone, which is uniformly colored due to uniformity of time averages in that region. 

Utilizing a single-function partition some independent invariant sets appear colored with the same color in some plots -- since the averaged functions are not one to one, same time average values can be obtained in dynamically different regions. This is why a single time average plot does not uniquely identify an ergodic set: this problem will be addressed later by employing more functions. Despite being obtainable with usual techniques, these results nicely illustrate the implementation of our method.

\subsection{Multi-scale methods}

Selecting different functions to time average  can reveal dynamics on different scales in the phase space. In general, more slowly-varying functions will reveal only the broad features of the phase space, while fast-variation and localized features can reveal smaller-scale dynamics.

\begin{figure*}[!hbt] \begin{center}
    \includegraphics[height=11.cm,width=12.5cm]{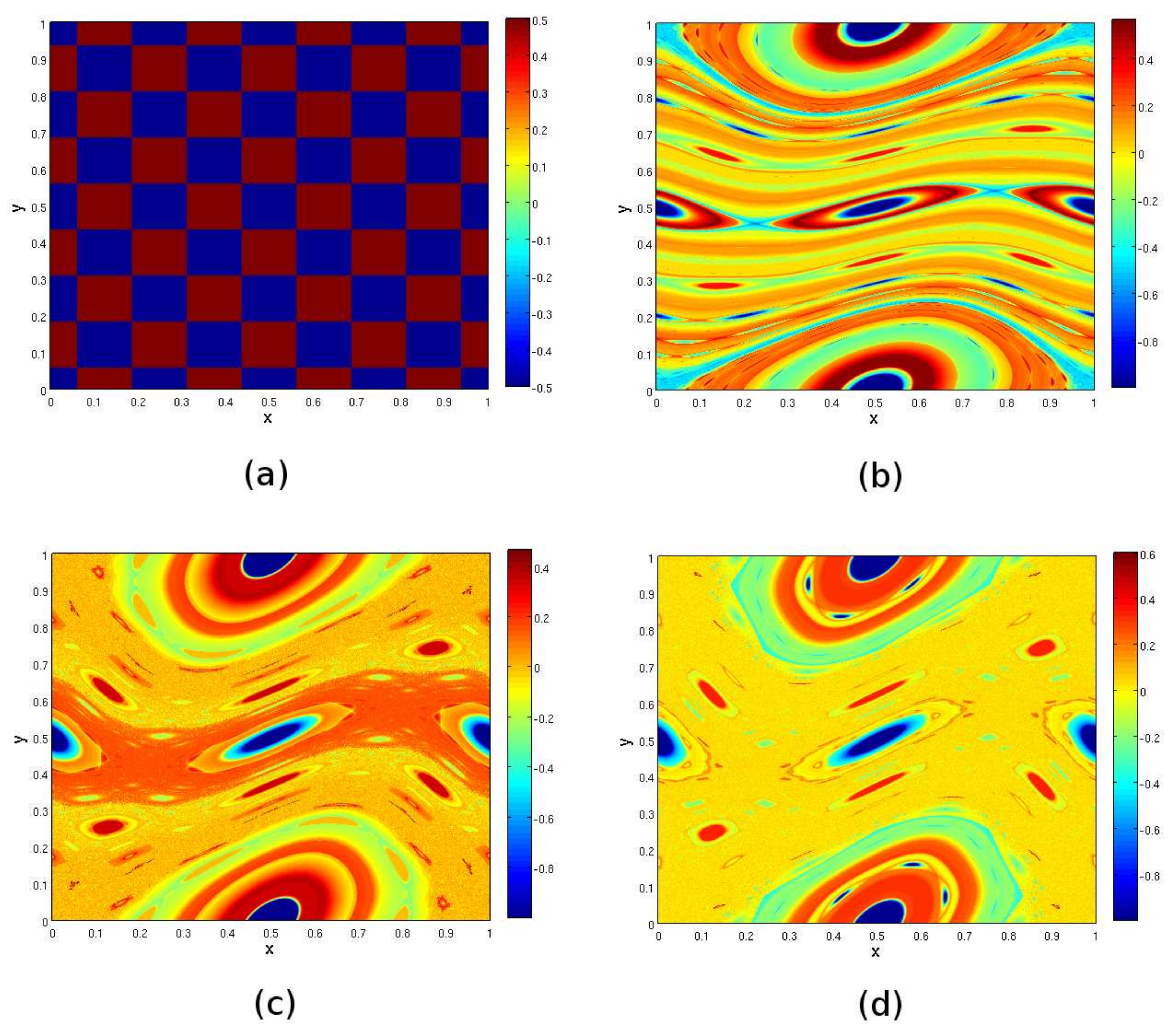}
\caption{The graph of the wavelet $W$ as defined by Eq.\,(\ref{waveletW}) in (a). Its time average under the dynamics of the standard map Eq.\,(\ref{mojasm}) for $\e=0.09$ in (b), $\e=0.16$ in (c) and $\e=0.18$ in (d). Grid: $800 \times 800$ and $t_{\mbox{final}}=50000$ iterations.} \label{wave}
\end{center}  \end{figure*}

As an example, we consider functions from the \textit{Haar wavelet} basis in 2D. Wavelets are a multi-scale functional family used in frequency decomposition and multi-resolution analysis \cite{kaiser}. In 1D, the Haar wavelet basis is constructed from the \textit{mother wavelet} $\psi_{00}$, defined by:
\[ \psi_{00} (x) = \left\{
\begin{array}{rll}
-1 & \mbox{if} & x \in  [0,\frac{1}{2}]  \\
1 & \mbox{if} & x \in \; ]\frac{1}{2},1[  \\
0 & \mbox{if} & x \in  \mathbb{R} \backslash [0,1]
\end{array} \right. \]
A general 1D wavelet $\psi_{ij}$ is constructed from this one by the appropriate transformations, summarized as:
\[ \psi_{ij} (x) := 2^{i/2} \psi_{00} (2^{i} x - 2^{-i} j) , \]
and,
as it can be shown, the functional family $\{ \psi_{ij} \}_{i,j \in \Z}$ is a basis for $L^2(\R)$. The 2D wavelets are constructed as products of 1D wavelets (in case 2D functional family is to be a basis for $L^2(\R^2)$, the procedure also involves cross-products of 1D wavelets with 1D \textit{scaling functions}; we shall however skip these technical details, and refer the reader to \cite{kaiser} and references therein). For the purposes of our study, we construct and use the 2D wavelet function called $W$:
\begin{equation}  \begin{array}{ll}
 W= & 2^{\frac{1}{3}} \sum_{n,m=1}^{8} \psi_{3n} \otimes \psi_{3m}  =  \\
 & = \sum_{n,m=1}^{8} \psi_{00} (8x-\frac{n}{8})\;\otimes\;\psi_{00} (8x-\frac{m}{8}),  \end{array} \label{waveletW}  \end{equation}
whose graph is shown in Fig.\,\ref{wave}a. Clearly, $W$ can be constructed from the 2D wavelets basis. Note that $W$ is not continuous
and thus does not strictly correspond to theory in \cite{mezic:1994}. However, this can be readily remedied using compactly supported continuous wavelets.

In Figs.\,\ref{wave}b,\,c\,\&\,d we show time averages of $W$ under the standard map dynamics  for various values of $\e$-values. It is again easy to recognize the known standard map features visible through diverse coloration of the phase space regions. As opposed to the Fourier functions case, wavelet functions have "sharp edges", which is why these plots have more drastic coloration changes among closely located \inv sets (compare Fig.\,\ref{wave}b to the middle plot in Fig.\,\ref{fourier} regarding $\e=0.09$, and Fig.\,\ref{wave}d to the last plot in Fig.\,\ref{fourier} regarding $\e=0.18$). This property can be used for "zooming" (see Section \ref{The Clustering Methods and Visualization}), e.g. in the case when a specific phase space sub-region is to be examined. It is interesting to note that the chaotic region for $\e=0.16$ (cf. Fig.\,\ref{wave}c) is not uniformly colored despite $\e$ being above the chaotic transition value - as the transport throughout the chaotic region is still very slow in the area around the last broken KAM curve, a different coloration occurs there. This type of chaotic transport in Hamiltonian maps can be approximated by the Markov tree model \cite{spiegel,cristadoro,meiss:trans}.

\begin{figure*}[!hbt] \begin{center}
    \includegraphics[height=14.cm,width=12.7cm]{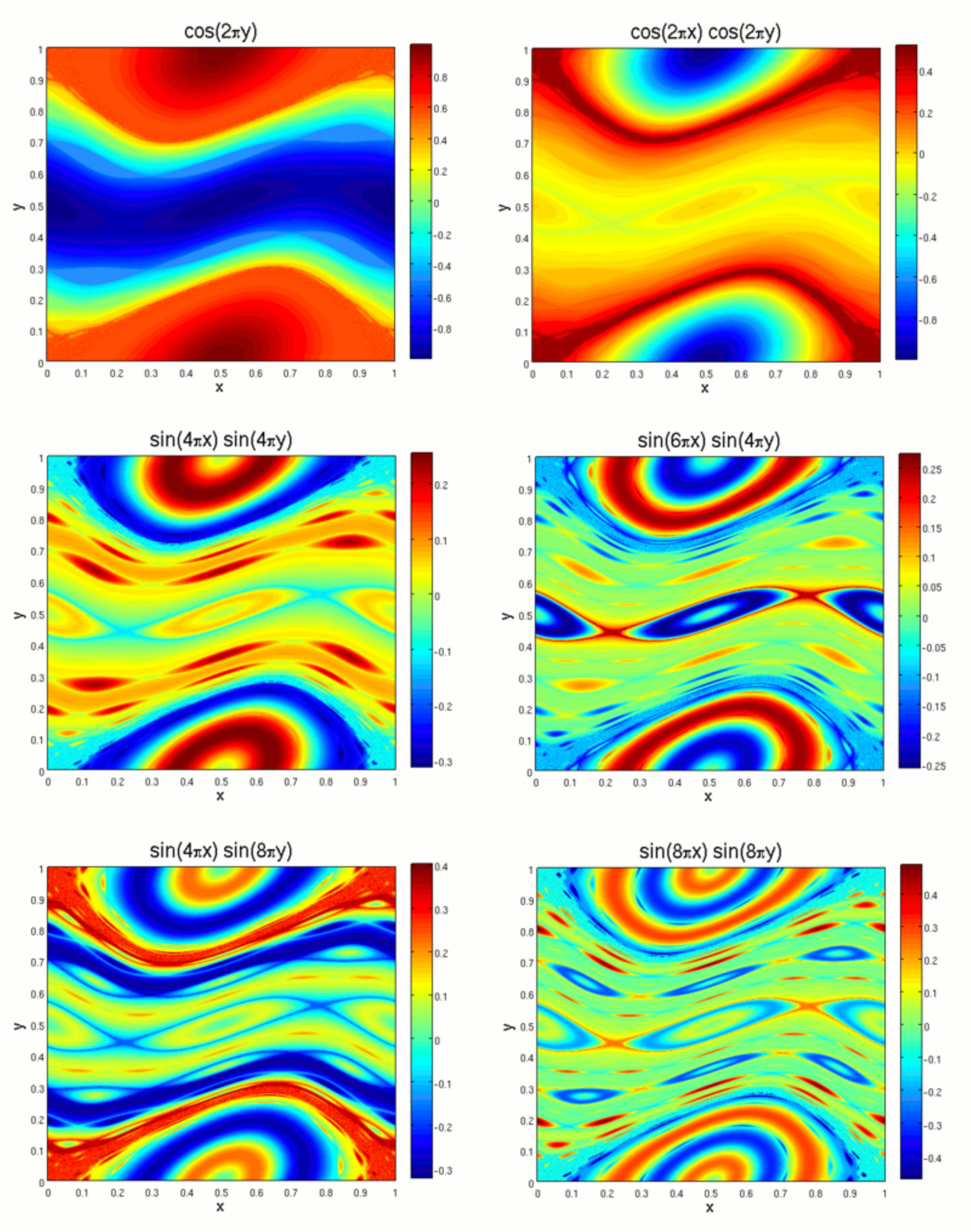}
    \caption{Color-plots for various time averaged functions for the standard map Eq.(\ref{mojasm}) with $\epsilon=0.12$. More detail of the dynamics (higher order resonant islands) 
       are revealed when the frequency of the function is increased. Grid: $800 \times 800$ and $t_{\mbox{final}}=30000$ iterations.} \label{fourscale}  
\end{center}  \end{figure*}

In order to investigate the phase space structure at larger (smaller) scale, one  takes smaller (bigger) Fourier/wavelet frequency. In the context of standard map it is convenient to use a slightly bigger frequency for $y$ coordinate - since the transport in $x$ direction is much faster, the function averages out to zero rather quickly in this direction. Time-average plots for several Fourier basis functions with increasing frequency are shown in Fig.\,\ref{fourscale}. Note the relationship between the used frequency and the scale of the best-visualized phase space details.

While the Fourier functions give smoothly colored plots, wavelets produce sharper and more detailed plots with a better distinction between the independent \inv sets. However, different \inv sets still happen to be assigned the same colors. To remedy this, we use the full power of the ergodic partition concept in Section \ref{The Clustering Methods and Visualization}.


\section{The Convergence Properties} \label{The Convergence Properties}

It is well known that time-averages of functions under dynamics of measure-preserving maps can converge arbitrarily slowly (see \cite{KachurovskiiKrengel} for discussion
and related results). However, for specific continuous functions bounds on rate of convergence are computable \cite{Avigadetal:2009}. Bounds that depend only on the 
maximum absolute value of a continuous function can be obtained  \cite{Avigadetal:2009}, but in our context it is relevant to study how convergence properties depend on 
the type of trajectory of $f$.
Thus, in this Section we study numerically the convergence properties of  time averages in order to estimate the precision of the values obtained and relate them to the different types of orbits. For simplicity we use only Fourier functions (the results for wavelets are similar). Consider the $t$-th partial time average for a function $f$ given by:
\begin{equation}
 f^t (x_0,y_0) = \frac{1}{t} \sum_{k=0}^{t-1} f({\T}^k (x_0,y_0)) ,
\end{equation}
with $\lim_{t \rightarrow \infty} f^t (x_0,y_0) = f^* (x_0,y_0)$ (which we assume exists for all grid-points $(x_0,y_0)$). The difference: 
\begin{equation}
 \Delta (t) = | f^{*}(x_0,y_0) - f^t (x_0,y_0) |  \label{conv-eq}
\end{equation}
is a sequence whose asymptotic behavior is to be studied in relation to the initial point $(x_0,y_0)$ and the $\e$-value, depending on the phase space regions with different dynamical behaviors. We consider the function $f=\cos(2\pi y)$ (cf. Fig.\,\ref{fourier}), define $f^*=f^{t}$ for $t=10^8$ and consider the first $10^6$ iterations.

\begin{figure*}[!hbt] \begin{center}
  \includegraphics[height=5.cm,width=16.cm]{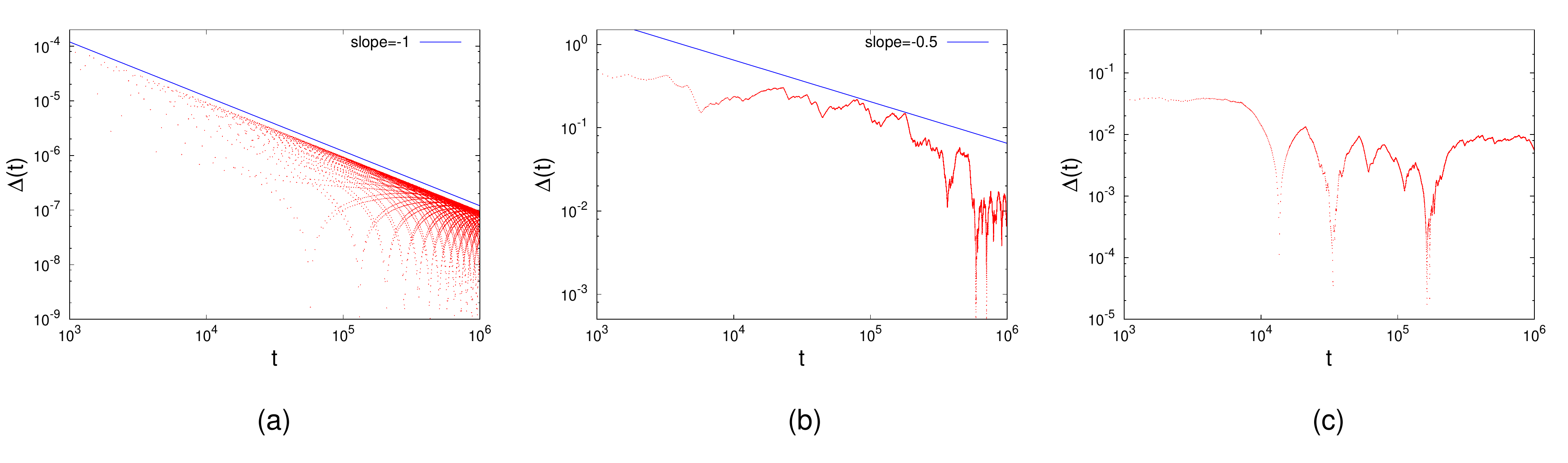} 
\caption{The convergence plots of the time averages for the standard map dynamics. The function $f=\cos(2\pi y)$ is considered, and $f^*$ is taken to be $f^{10^8}$. (a):
a regular orbit for $\e = 0.09$ (cf. Fig.\,\ref{fourier}) for the initial point $(0.5,0.4)$ fitted with the slope of -1; (b): a strongly chaotic orbit for $\e = 0.18$ (cf. Fig.\,\ref{fourier}) for the initial point $(0.02,0.02)$ fitted with the slope of $-\frac{1}{2}$; (c) a weakly chaotic orbit for $\e = 0.09$ (cf. Fig.\,\ref{fourier}) for the initial point $(0.01,0.01)$.}  \label{convergence} 
\end{center}  \end{figure*}

\textit{The Regular Region}. For all the points on the regular trajectories the time averages of continuous functions converge with the error decreasing as $\frac{a}{t}$, with the constant $a$ given by the trajectory properties, as shown in \cite{mezicsoti}. The result applies uniformly to all the regular (periodic or quasi-periodic) orbits, regardless of the $\e$-value and the choice of function. Given that $a \sim O(1)$ this allows a rather precise estimation of the final precision of the obtained time average value, in relation to the total number of iterations computed $t_{\mbox{final}}$. A typical convergence plot for the case of a regular trajectory is reported in Fig.\,\ref{convergence}a.

\textit{Strongly Chaotic Region}. In the case of strong mixing the fluctuations of the time average decrease as $\frac{1}{\sqrt{t}}$ \cite{mezicsoti}.  Our findings  indicate this rate asymptotically approaches $t^{\alpha}$ regime with $\alpha \gtrsim -\frac{1}{2}$; the $\alpha=-\frac{1}{2}$ result was obtained only in the case of very chaotic orbits (large $\e$). A typical convergence plot for a chaotic trajectory is shown in Fig.\,\ref{convergence}b. Given the mixed phase space of map Eq.\,(\ref{mojasm}), the plot exhibits a very irregular behavior, which is bounded by the $t^{\alpha}$ regime away from the transients. Hence, the error in the case of chaotic time averages can be estimated from below by  $1/\sqrt{t_{\mbox{final}}}$ and improved by the increase of the $\e$-value.

\textit{Weakly Chaotic Region}. Below the chaotic transition for Eq.\,(\ref{mojasm}) ($\e \lesssim 0.15$), the chaotic regions are localized in the phase space around the hyperbolic fixed points, and characterized by a weak chaos with trajectories slowly diffusing through the region \cite{meiss:trans}. The convergence of the time averages in this region is extremely slow and irregular, and  cannot be in general bounded by any asymptotic slope of $t^{\alpha}$ type. A typical convergence plot for this region is reported in Fig.\,\ref{convergence}c. Note that despite plot showing less irregular oscillations than in the strongly chaotic case, it barely decreases and cannot be fitted with a determined slope. This convergence pattern is consistently present in all the weak chaos trajectories, and it improves only with the increase of $\e$-values. It is therefore hard to estimate the final error in this case, unless the dynamics is run for excessively long times. 

Our visualization method appears to suit better the cases of either regular or strongly chaotic behavior where the ``speed'' of filling the \inv set is relatively high. However, we find that even in the case when weak chaotic behavior is present, we obtain a good representation of the structure of the phase space, given that regular and strongly chaotic orbits fill a large portion of the phase space. In view of this, one can set the total number of iterations according to the precision rates of the strongly chaotic zone. We thus typically take for standard map $t_{\mbox{final}} \sim O(10^4)$ iterations: this sets the precision of strongly chaotic case to $O(10^{-2})$ (with even better precision for the regular case), which is enough given that the considered functions have values in the $[-1,1]$ interval. Note also that the convergence properties show a certain pattern in relation to the trajectory type: it would be possible to characterize the nature of a trajectory by looking at this patterns for different functions (similarly as done in \cite{gottwald}). In Figs.\,\ref{clustershrinking}\,\&\,\ref{esm} we examine the convergence slopes for more functions, and averaged over and ensemble of trajectories.


\section{Ergodic Quotient Space and Clustering Methods} \label{The Clustering Methods and Visualization}

In this Section we present the concept of the {\it ergodic quotient space} and an algorithm based on time averages of functions that helps us understand its topology. 

A discrete-time measure-preserving map $\x' = \T \x$ on a compact phase space $A$ decomposes the subset of the phase space $\Sigma$ on which time averages of all continuous functions exist  into ergodic sets \cite{mezic:1994,mezicwiggy}. The ergodic quotient space $Q_e$ is the space where each ergodic set is mapped into a single point, and, additionally, the complement of $\Sigma$ is mapped into a point. Consider the space $S$ of all the infinite sequences indexed by non-negative integers $S=\{a_0,a_1,a_2,...\}$. Consider also a basis ${f_i}, i\in \mathbb{N}$ for $L^2$ on $A$. Denote the image of $q:\Sigma\rightarrow S, q(\x)=\{\chi_{\Sigma^c}(\x),f_1^*(\x),f_2^*(\x),...\}$ by $Q_e^*$, where $\chi_{\Sigma^c}$ is the indicator function on set $\Sigma^c$ and call the map $q$ the ergodic quotient map. For convenience we extend the definition of $f_i^*$ from $\Sigma$ to the whole set $A$ by setting $f_i^*=0$ on $\Sigma^c$. Then, clearly, $$Q_e=Q_e^*\cup\{1,0,0,...\}.$$ Now note that finite-dimensional Euclidean space embeddings of projections of $Q_e$ to a space of finite sequences labeled by $i_1,...,i_N$ can be obtained by $N$-tuples $\{f_{i_1}^*,...,f_{i_N}^*\}(\x)$. We call  such embeddings {\it Mesochronic Scatter Plots} and study them numerically in this Section. But before that, let us give an example  for ergodic quotient space of the discrete dynamical system presented in Eq.\,(\ref{mojasm}) with $\epsilon=0$.  In that case, fixing an irrational $y$, the time average of any function is constant, for any initial $x$. Thus, the whole circle $y=const.$ is mapped into a single point. For rational values of $y=\frac{p}{q}$ ($\frac{p}{q}$ is an irreducible fraction), where the invariant circle at that $y$ is filled with periodic orbits, the length of the interval parameterizing distinct periodic orbits is $\frac{1}{q}$. It is then easy to see that $Q_e$ is a "rational comb", consisting of a straight line with an interval of length $\frac{1}{q}$ attached at every rational $y=\frac{p}{q}$.

\subsection{2-dimensional MSP embedding}

To provide a graphical representation of $Q_e$, we consider Fourier basis functions and begin by the  case of $N=2$ basis functions. Time averages $f^*_1$ and $f^*_3$ are computed for every grid-point defining the correspondence between the grid-points $(x_0,y_0)$ and the time average vector $\{f^*_1,f^*_3\}$:
\[ (x_0,y_0) \in \mbox{grid} \;\;\; \longrightarrow \;\;\; (f^*_1(x_0,y_0),f^*_3(x_0,y_0)) \in [-1,1]^2 . \] 
The MSP is obtained by plotting all the vectors $(f^*_1,f^*_3)$ on the square $[-1,1]^2$.  We use the grid of  $300 \times 300$ points and run the dynamics for $30000$ iterations.

\begin{figure*}[!hbt] \begin{center} 
    \includegraphics[height=12.cm,width=14.5cm]{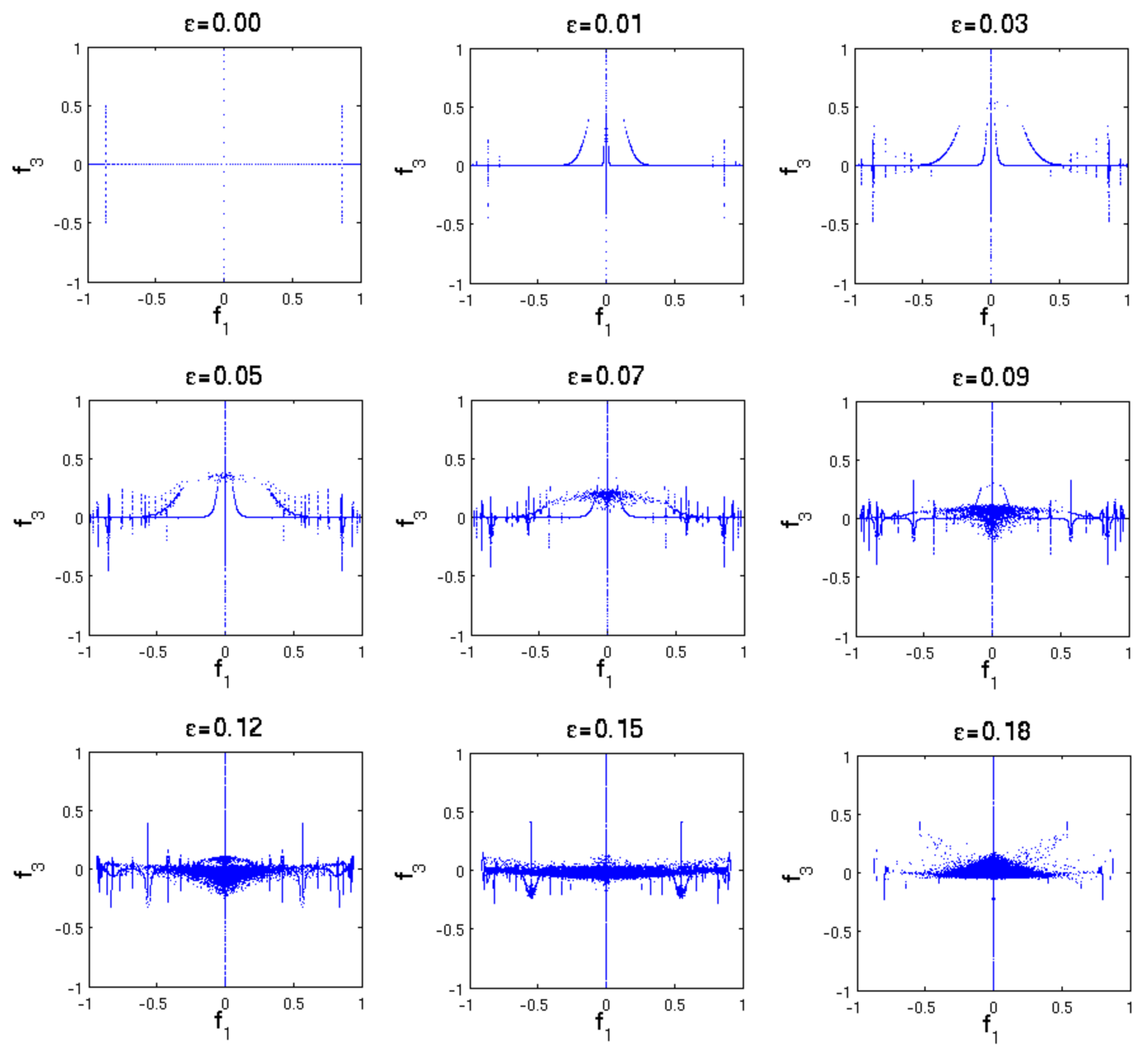}
 \caption{Sequence of 2D MSPs for standard map Eq.\,(\ref{mojasm}), with functions $f_1=\sin(2\pi y), f_3=\cos(12\pi x)\cos(2\pi y)$. The value of $\e$ is indicated in each plot. Each 
time average was obtained on a $300 \times 300$ grid, for $30000$ iterations.}  \label{sequence2d}  
\end{center}  \end{figure*}

In Fig.\,\ref{sequence2d} we show a sequence of 2D MSP for the functions $f_1=\sin(2\pi y)$ and  $f_3=\cos(12\pi x)\cos(2\pi y)$ with increasing $\e$-value. In the first plot showing the case of $\epsilon=0$, for irrational $y$, the time average of $f_1$ is $\sin(2\pi y)$, while the time average of $f_3$ is $0$. This explains the horizontal line in the $\epsilon=0$ plot. For rational $y$ the time average of $f_3$ is only non-zero provided $y=0,\frac{1}{2},\frac{1}{3},\frac{2}{3},\frac{1}{6}$ and $y=\frac{5}{6}$ (in other words, the function $f_3$ "resonates" with the map dynamics and produces non-zero values of time averages only for those $y's$). These values contribute the vertical lines in the $\epsilon=0$ plot. However, some of these vertical lines overlap in this projection, which will lead us to consider three-dimensional projections later. The side lines will become more evident with increasing $\epsilon$, due to the appearance of resonance zones and the associated new families of quasiperiodic orbits. With further increase of $\e$, the side lines disappear and a central scattered region appears, leading to the chaotic transition at the familiar value close to $\e \thickapprox 0.154$. 

\begin{figure*}[!hbt] \begin{center} 
\includegraphics[height=8.cm,width=13.cm]{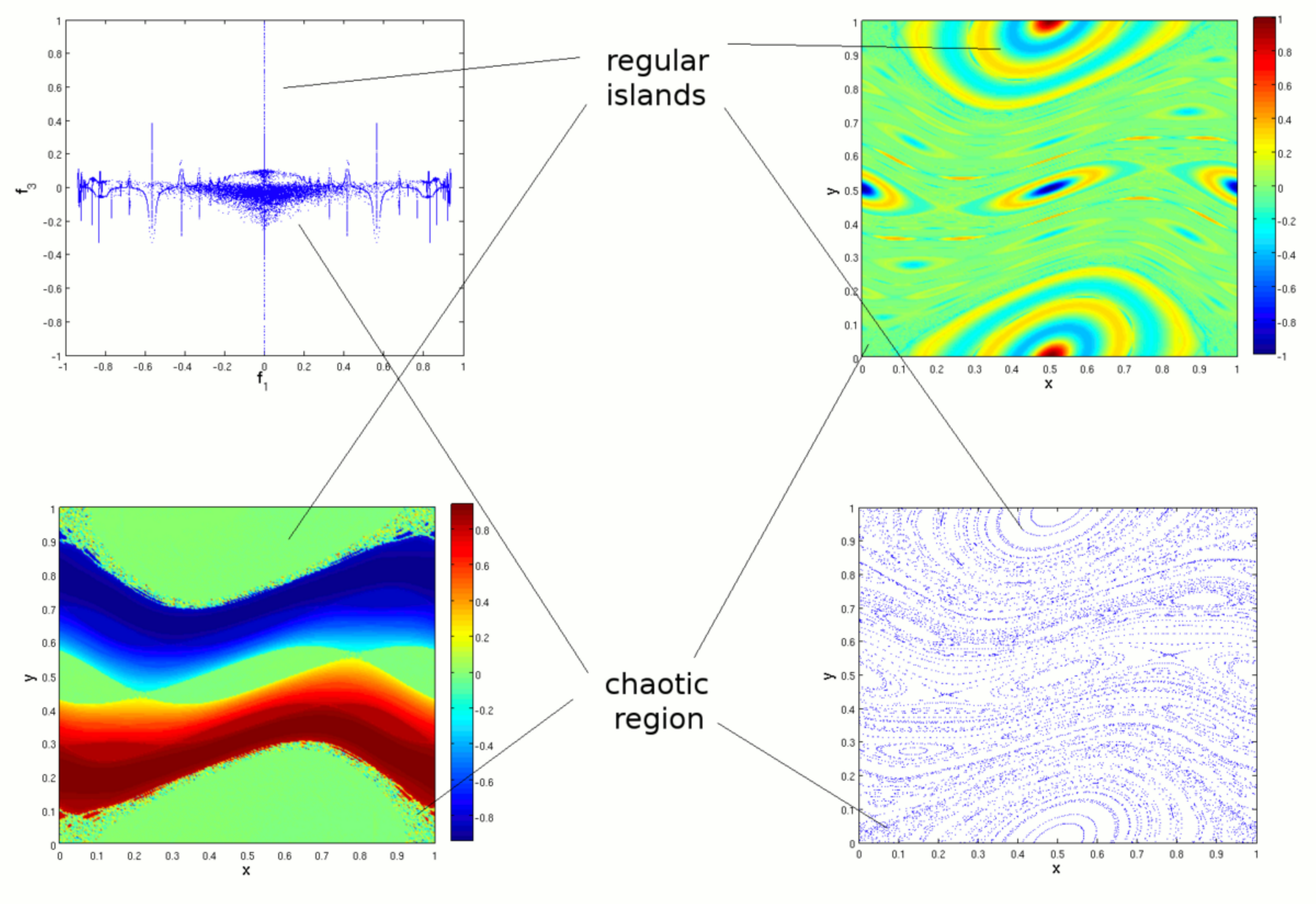}
\caption{Two-function MSP  (up left) for the standard map Eq.\,(\ref{mojasm}) for $\e=0.12$ (cf. Fig.\,\ref{sequence2d}). Time-averaged functions are $f_1=\sin(2\pi y)$ (down left) and $f_3=\cos(12\pi x)\cos(2\pi y)$ (up right). Phase space dynamical regions are indicated, in correspondence with the MSP parts and the dynamical regions in phase space portrait (down right).}   \label{scatter-e012} 
\end{center} \end{figure*}

The idea is further illustrated in Fig.\,\ref{scatter-e012} where a MSP  involving two functions is investigated with reference to the corresponding time averages plots and the phase space portrait. As already stated, each single point in the MSP has the $x$-coordinate equal to $f_1^*(x_0,y_0)$ and the $y$-coordinate equal to $f_3^*(x_0,y_0)$ for some grid-point $(x_0,y_0)$. As indicated in the figure, long branches (curves) represent the families of periodic islands around elliptic fixed points, while the irregular clouds amount for localized chaos around the hyperbolic fixed points. Secondary chaotic zones appearing around second-order hyperbolic points are also visible, together with the secondary families of periodic orbits.

\begin{figure*}[!hbt] \begin{center} 
\includegraphics[height=8.cm,width=13.cm]{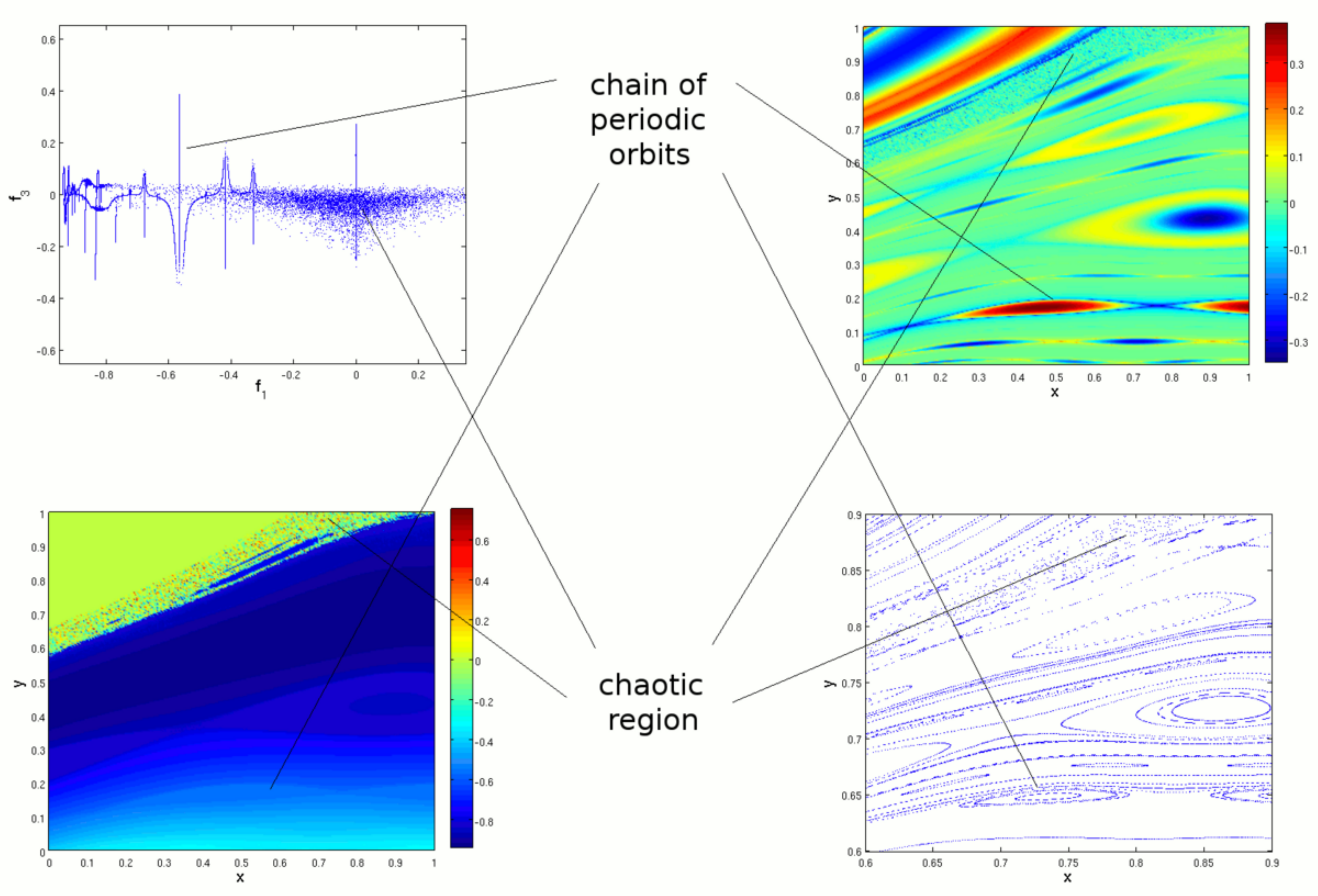}
\caption{A zoomed part of the MSP (up left) from Fig.\,\ref{scatter-e012} for the phase space region $[0.6,0.9]\times [0.6,0.9]$ (for $\e=0.12$, $f_1=\sin(2\pi y)$ (down left) and $f_3=\cos(12\pi x)\cos(2\pi y)$ (up right)). Dynamical regions and their corresponding MSP parts are indicated are related to the phase space portrait (down right).} \label{scatter-e012-zoom}  
\end{center} \end{figure*}

To investigate this further we show the same MSP for the same time averages, zoomed to the phase space region $[0.6,0.9]\times [0.6,0.9]$ in Fig.\,\ref{scatter-e012-zoom}. This MSP can readily be recognized as a part of the MSP from Fig.\,\ref{scatter-e012}. Again, we see the interplay between long curved lines and irregular clouds, representing regular and chaotic regions respectively. Note that many more secondary periodic families are visible in this plot due to the improved resolution (zoom), capturing the scale-invariant fractal nature of the standard map's phase space.

\subsection{3-dimensional MSP embedding}

The two-dimensional projections presented in the previous Section have the unpleasant feature of self-intersection. Three-dimensional embeddings resolve this issue for two dimensional maps. We will see later that higher dimensional embeddings are needed to avoid intersections in higher dimensional maps. 

\begin{figure*}[!ht] \begin{center} 
    \includegraphics[height=12.cm,width=14.5cm]{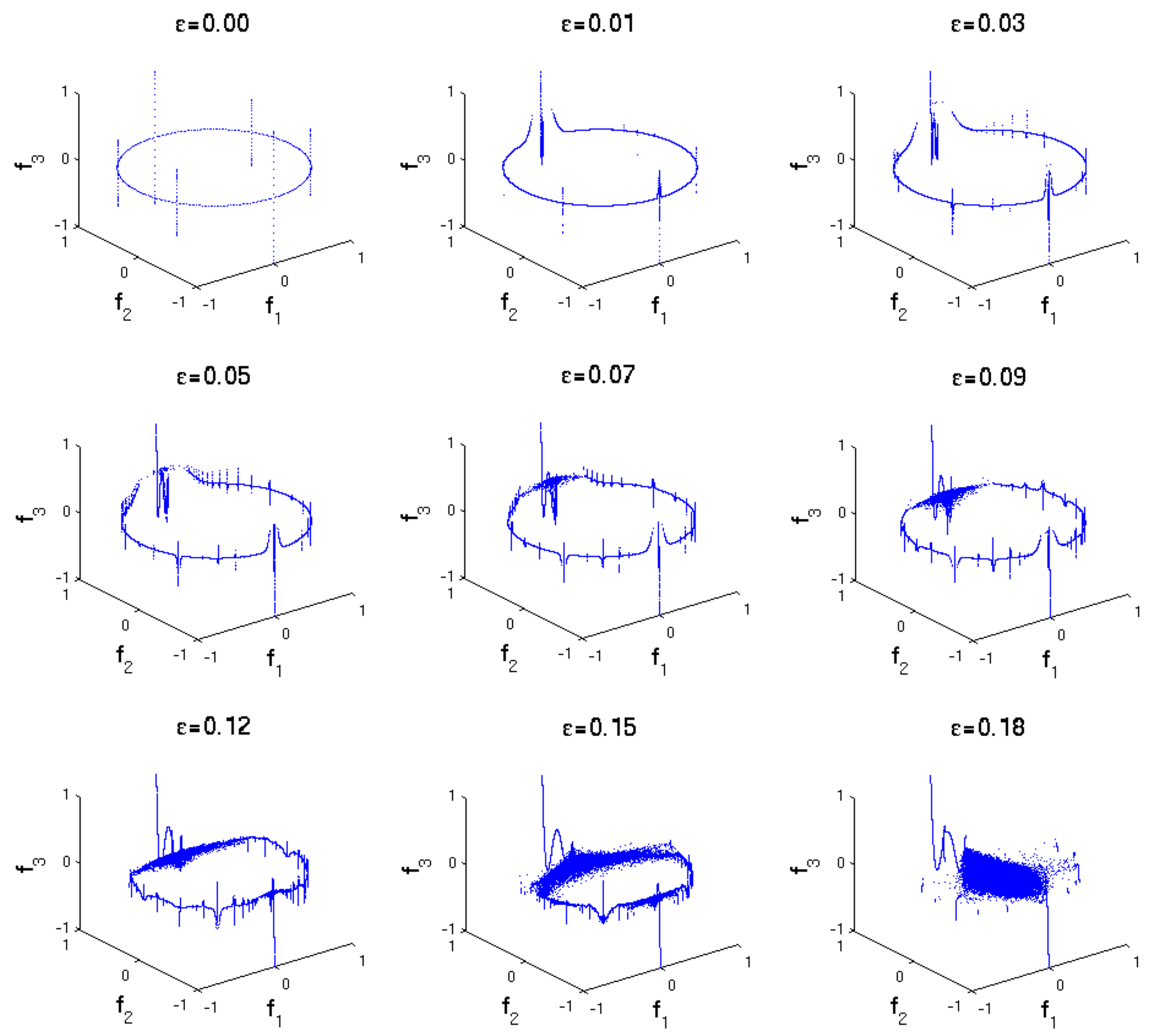}
 \caption{Sequence of 3D MSPs for the standard map Eq.\,(\ref{mojasm}), with functions $f_1=\sin(2\pi y)$, $f_2=\cos(2\pi y)$ and  $f_3=\cos(12\pi x)\cos(2\pi y)$. The value of $\e$ 
is indicated in each plot. Each time average was obtained on a $300 \times 300$ grid, for $30000$ iterations.}  \label{sequence3d}  
\end{center}  \end{figure*}

We set $N=3$ and consider the MSPs done with three linearly independent functions.  Using the same grid and total iterations as previously, we consider the correspondence:  
\[ (x_0,y_0)  \longrightarrow (f^*_1(x_0,y_0),f^*_2(x_0,y_0),f^*_3(x_0,y_0) ) \in [-1,1]^3 , \] 
with $f_2=\cos(2\pi y)$. In Fig.\,\ref{sequence3d} we show nine MSPs obtained for the three functions and increasing $\e$-values. We monitor the phase space structure evolution as $\e$ is changed, in terms of geometric complexity evolution of the MSP's structure, which is much easier to do within this 3D embedding. The size of resonance zone emanating from circle of period-6 orbits is substantially shrunk compared with that for period-3, period-2 and period-1, even for small, $\e=0.01$ perturbation (middle figure of top row in Fig.\,\ref{sequence3d}).  Note the changes in branches and development of higher order periodic islands, with increased $\e$ with localized chaotic zones around hyperbolic periodic orbits. These secondary islands have resonance zones of their own that merge and enable the chaotic transition, where a single large chaotic zone enables trajectories to pass around the torus in both directions. The chaotic regions are visible as thickened scatter in the plots. The chaotic transition occurring for $\e \thickapprox 0.154$ can here be seen as a merging of localized chaotic zones that propagate along the branches into a single connected chaotic zone, visible as a single cluster for the value $\e=0.18$ in Fig.\,\ref{sequence3d}. Also visible in the $\epsilon=0.18$ plot are the side lines corresponding to the remaining islands around $y=0,\frac{1}{2},\frac{1}{3}$ (see bottom plot in Fig.\,\ref{fourier}). As expected, in the strongly chaotic regime all the time average vectors are localized in a single cluster that shrinks in size with further increase of $\e$. In the case of ergodic behavior one expects all the time average vectors to shrink to  a single point in $[-1,1]^3$ space. 

Adding a function to the MSP, and thus increasing the embedding dimension  clearly improves the representation of the phase space structure. Furthermore, the Fig.\,\ref{sequence3d} shows  the sufficient embedding dimension for the standard map's MSPs to be three: it is in three dimensions where the families of regular orbits can be fully represented without intersection.

\begin{figure*}[!hbt] \begin{center}
        \includegraphics[height=6.cm,width=14.cm]{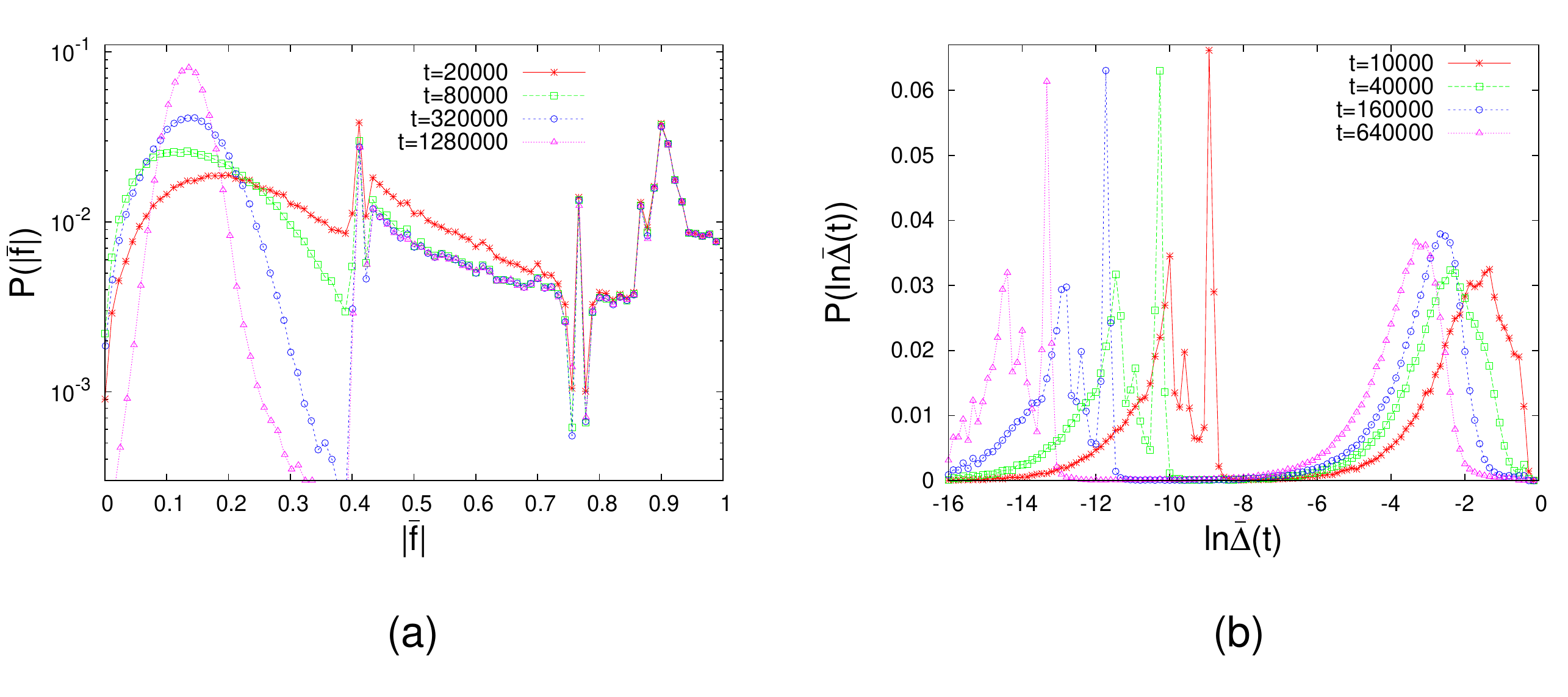}
\caption{Time averages for the functions $f_1=\sin(2\pi y)$, $f_2=\cos(2\pi y)$ and $f_3=\cos(12\pi x)\cos(2\pi y)$ are computed for various iteration-values for the standard map Eq.\,(\ref{mojasm}) on the grid of 
$400 \times 400$ with $\e=0.18$ (cf. last plot in Fig.\,\ref{sequence3d}), and the time-evolution of the norm of time average vectors $|\mathbf{\bar f}^t|=\sqrt{(f_1^t)^2 + (f_2^t)^2 + (f_3^t)^2}$ is considered. (a): distributions of $|\mathbf{\bar f}^t|$ as function of time; (b): distributions of ${\bar  \Delta}(t) = \big| |\mathbf{\bar f}^*| - |\mathbf{\bar f}^t| \big|$ as function of time, done by taking  
$|\mathbf{\bar f}^*| = |\mathbf{\bar f}^{t=1.28 \times 10^{6}}|$.}\label{clustershrinking}
\end{center} \end{figure*}

Finally, in Fig.\,\ref{clustershrinking} we examine the time-evolution of the time average vectors for $\e=0.18$, shown in the last plot in Fig.\,\ref{sequence3d}. Time averages of the same functions are computed for various final iteration-values $t_{\mbox{final}}$, and for each grid point $(x_0,y_0)$ the norm of time average vector
\begin{equation}
  |\mathbf{\bar f}^t (x_0,y_0) |=\sqrt{(f_1^t (x_0,y_0))^2  + (f_2^t (x_0,y_0))^2 + (f_3^t (x_0,y_0))^2 }   \label{tav-norm}
\end{equation}
is considered. The distribution of values of $|\mathbf{\bar f}^t|$ is shown in Fig.\,\ref{clustershrinking}a as function of time $t$.  While for large $|\mathbf{\bar f}^t|$-values we see a quick convergence into a final profile, for small $|\mathbf{\bar f}^t|$-values the distribution-profile seems to be slowly evolving towards a sharp single-peaked (delta) distribution. Clearly, the former corresponds to the regular orbits that are faster to converge to final $f^*$-values, while the latter corresponds to the slowly converging chaotic orbits (cf. Section \ref{The Convergence Properties}). We illustrate this further be generalizing the Eq.\,\ref{conv-eq} for the case of more functions into:
\begin{equation}
{\bar  \Delta}(t)  =  \big| |\mathbf{\bar f}^{*} (x_0,y_0) |  -   |\mathbf{\bar f}^{t} (x_0,y_0)| \big|  
\end{equation}
Thus, ${\bar  \Delta}(t)$ measures how close is the norm of the partial time average vector $|\mathbf{\bar f}^{t}|$ to its limit value $|\mathbf{\bar f}^{*}|$. In Fig.\,\ref{clustershrinking}b we show the distribution of ${\bar  \Delta}(t)$ for all grid-points in function of time $t$, obtained by taking $|\mathbf{\bar f}^*| = |\mathbf{\bar f}^{t=1.28 \times 10^{6}}|$. Two groups of peaks that travel to zero ($\ln {\bar  \Delta}(t) \rightarrow -\infty)$ and statistically maintain their shapes can be readily recognized. The structured group consisting of few smaller peaks has much smaller  $\ln {\bar  \Delta}(t)$-values than the single structureless peak located around $\ln {\bar  \Delta}(t) \sim -3$. Also, the former group of peaks seems to travel about twice faster towards zero than the latter single peak. This again corresponds to the difference between regular orbits (structured peak group) and the chaotic orbits (the single structureless peak) -- the former converge to zero with a rate of $t^{-1}$ uniformly for every point, while the latter \textit{on average} converge to zero with the rate of $t^{-0.5}$. 

\subsection{Visualization of the ergodic partition via clustering}

\begin{figure*}[!hbt] \begin{center}
      \includegraphics[height=11.5cm,width=12.5cm]{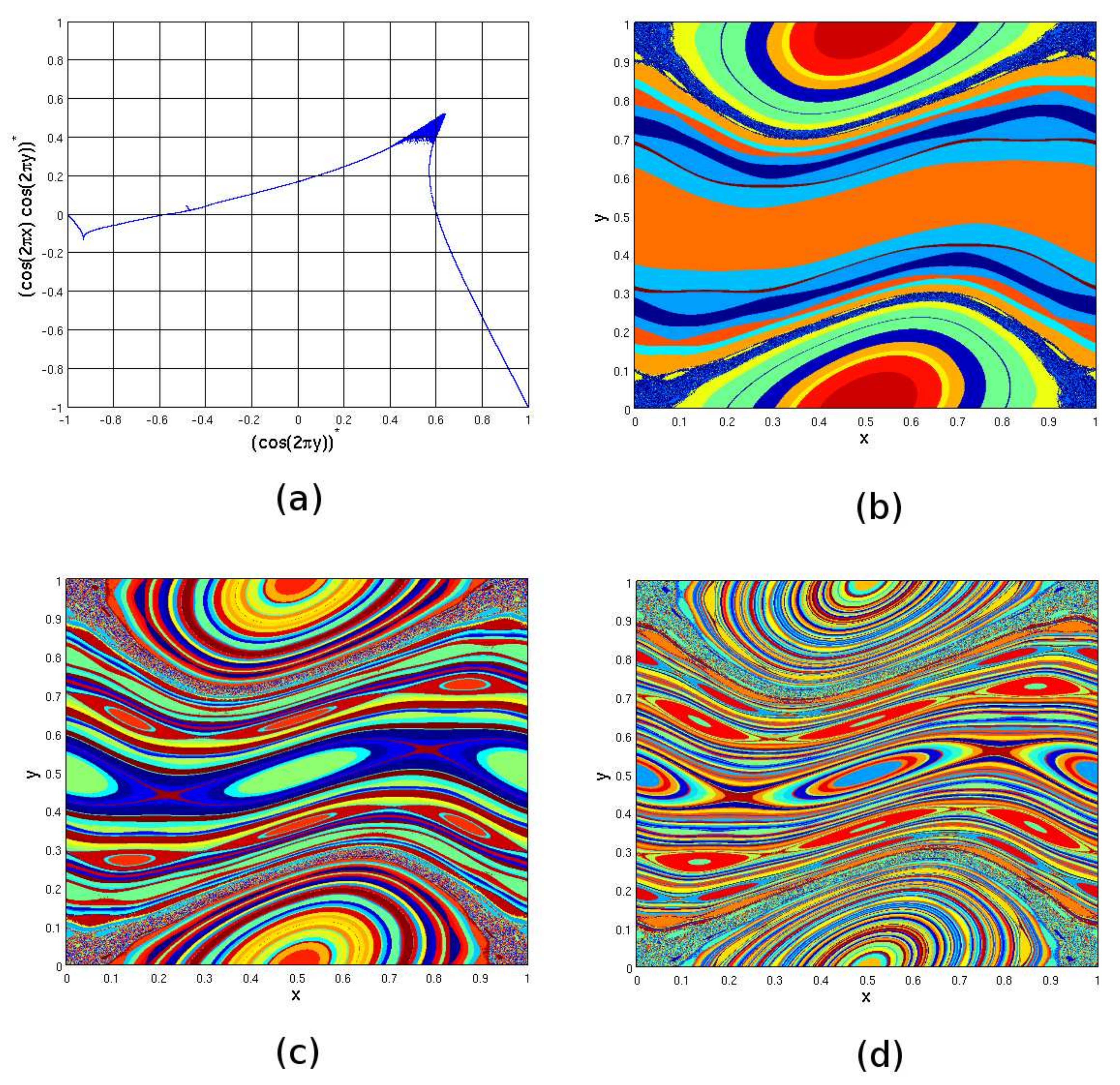}
\caption{Two-function approximation of the ergodic partition. (a): two-function MSP for the standard map Eq.\,(\ref{mojasm}) for $\e=0.12$, done using $\cos (2 \pi y)$ and 
$\cos (2 \pi x) \cos(2 \pi y)$, for a grid $800 \times 800$ and with $t_{\mbox{final}}=30000$. Three approximations of the ergodic partition done for different values of  
cell division $L$ (and constructed using this MSP for clustering), are shown in (b) for $L=10$, in (c) for $L=50$, and in (d) for $L=140$.}  \label{clust-1}
\end{center}   \end{figure*}

Following the investigation of the MSPs we construct a simple algorithm for approximation of the ergodic partition and its graphical phase space visualization.  Consider an $N$-function MSP contained in $[-1,1]^N$:

\begin{description}
\item[step 1] divide the $N$-cube $[-1,1]^N$ into $L^N$ cells dividing each axis into $L$ segments as illustrated in Fig.\,\ref{clust-1}a (for $N=2$ and $L=10$), and consider the distribution of time average vectors around the cells
\item[step 2] disregard the cells that contain no time average vectors
\item[step 3] assign a color to every remaining cell, therefore  assigning a color to every time average vector
\item[step 4] observe the grid-points corresponding to time average vectors sharing a cell/color: they define an $N$-order approximation of an ergodic set
\item[step 5] color the phase space by coloring each grid-point with the color assigned to it
\end{description}
Note that this is a generalization of the single-function coloring scheme with a difference that now the scheme is regulated by adjusting the cell division and optimizing it according to the structure of the MSP. The multi-dimensionality of the color-assigning rule  allows for higher differentiation among the \inv sets. 

For simplicity we start again with the case of two functions: one-dimensional lines intersect only at points and this feature does not  appear to perturb the phase space representation much.
We examine the two-function MSP showing it in Fig.\,\ref{clust-1}a with $10 \times 10$ cells division. The corresponding approximation of the ergodic partition with colors between blue and red randomly and uniformly assigned to non-empty cells is shown in Fig.\,\ref{clust-1}b. A better overall clarity of the \inv set structure is obtained both at the global and local level (within the approximation precision, which is also influenced by a limited number of available colors). Note that the visibility can be enhanced by coloring nearby \inv sets with different colors which is attained by randomizing the color-assignment for the non-empty cells. The number of visualized sets increases with increase of $L$ (Figs.\,\ref{clust-1}b\,\&\,c), but the color differentiation gets poorer, as the number of available (visible) colors remains limited (not only by the software, but also by the human eye recognition). The optimal value of $L$ (regulating the number of cells) is to be set according to the visualization requirements, taking into account the relationship between color differentiation vs. number of visible \inv sets. For the two-function case examined in Fig.\,\ref{clust-1}, it appears the optimal $L$ is around 50 (corresponding to Fig.\,\ref{clust-1}c). Too small $L$ is underusing  the MSP as it colors too many different \inv sets uniformly (Fig.\,\ref{clust-1}b), while for too large values of $L$ the lack of colors  brings the same problem, in addition to a very non-uniform coloration of the chaotic region (Fig.\,\ref{clust-1}d).

\begin{figure*}[!hbt] \begin{center}
       \includegraphics[height=5.cm,width=12.5cm]{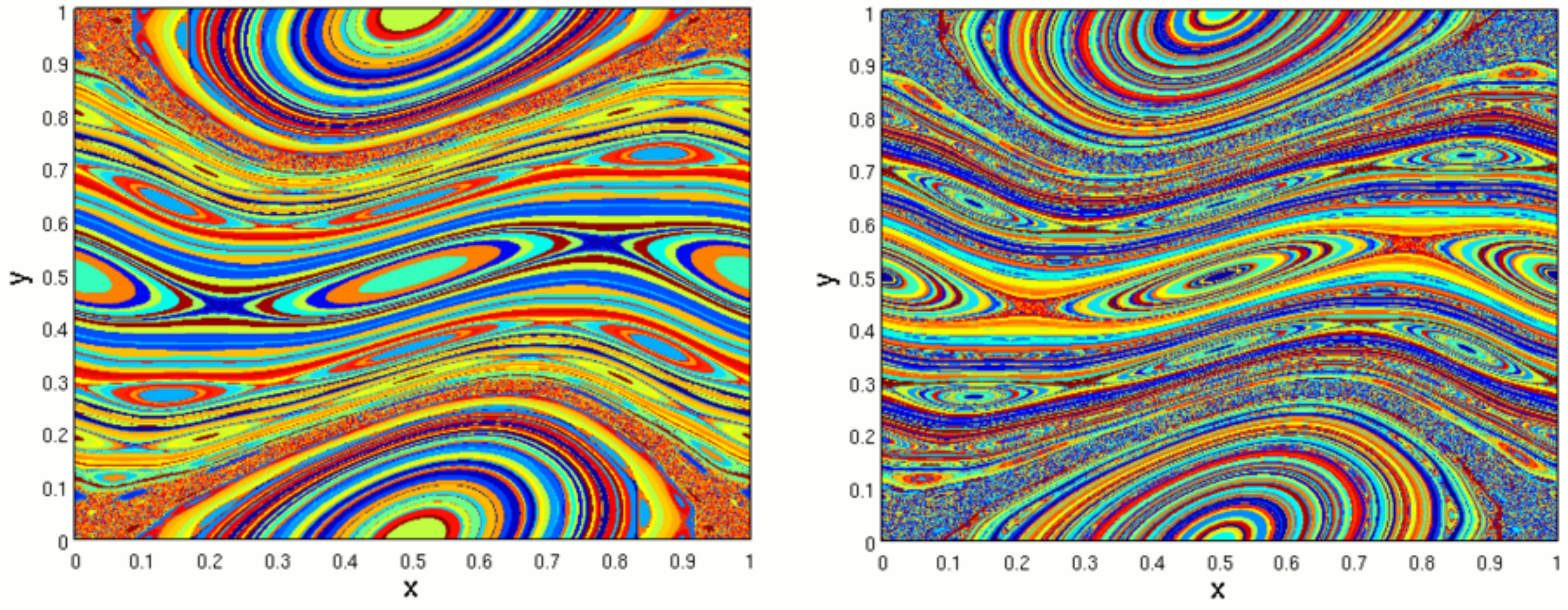}
\caption{Approximations of the ergodic partition for the standard map Eq.\,(\ref{mojasm}) with $\e=0.12$. Left: three-function approximation using two function from Fig.\,\ref{clust-1} and the function $\sin(4 \pi x) \sin(4 \pi y)$. Right: four-function approximation using these three functions in addition to $\sin(10 \pi x) \sin(10 \pi y)$. The grid  $800 \times 800$ is used for all time averages and the cell division for both approximations is $L=50$.} \label{clust-2} \end{center}  \end{figure*}

In Fig.\,\ref{clust-2} on the left we show an example of ergodic partition approximation constructed from a three-function MSP, using the functions from Fig.\,\ref{clust-1} case as the first two. Note that for $L=50$ we obtain a better quality than previously. Finally, in Fig.\,\ref{clust-2} on the right we add another function and show a four-function approximation obtained for the optimal $L$ value of $L=50$. 

As noted earlier, we have chosen functions involving different frequencies, thus visualizing global and local phase space features simultaneously. Fig.\,\ref{clust-2} reveals high-resolution approximations to the ergodic partitions for $\e=0.12$, producing good approximations to the \inv set structure of the standard map's phase space. Note that both pictures indeed visualize details at all scales, with the plot on the right being somewhat sharper.  In construction of these plots we sought to improve the cell division by having a relatively uniform number of time average vectors within each cell, in relation to the available colors. Due to a particular choice of functions, in the right plot on Fig.\,\ref{clust-2} we managed to obtain good coloration for $L=50$ (very large $L$-value considering the number of functions involved), creating the optimal approximation for the four-function case. Given the number of \inv sets visualized, different realizations of random colors assignments make very little difference in the overall picture.

The limited number of available colors still makes some different ergodic sets appear in the same color; this problem can be partially overcome by optimizing the color-assigning rule. Instead of assigning a random color to each non-empty cell one could create an assigning algorithm that would be optimized in relation to the particular case studied. Another problem arises in relation to the ergodic zone in the phase space: given that its time average vectors are more diffused then the regular orbits' ones (cf. Fig.\,\ref{sequence3d}), they set the lower bound to the size of the cells. This is why the ergodic zone in the Fig.\,\ref{clust-1}d appears non-uniformly colored. In the context of the standard map, it is convenient to pick a smaller $L$ for low $\e$-values in order to obtain a better focus on the nested invariant curves, while for larger $\e$-values a bigger $L$ allows to include the whole chaotic zone in a single cell/color.  Also, a bigger number of functions allows more flexibility for the $L$-value as the underlying MSP differentiates better among the \inv sets. Moreover, a better cell division scheme would not employ the simple cubical division described here, but a more sophisticated algorithm optimizing (for instance) the number of time average vectors per cell with respect to their distance in the time average space. One could also seek to adjust the shape of cell according to the properties of the MSP instead of just using the simple cubical ones. Furthermore, the problem of limited number of colors could be tackled by employing a specific graphically-oriented visualization software allowing more flexibility in terms of choosing or adjusting colors and their tones. Also, a cell's color could be determined in relation to the number of time average vectors contained in it, or contained in the neighboring cells, thus differentiating better among various \inv sets.

Finally, in the case of measure-preserving maps with a given maximum MSP embedding dimension (three in this case, cf. Fig.\,\ref{sequence3d}), one  could seek to parameterize the MSP obtaining a continuous coloration scheme that would include all the \inv sets. Still, this procedure would yield a non-uniform number of time average vectors per cell, allowing for further improvements. For the purpose of this study however, we limit ourselves to the simple and illustrative algorithm just exposed.


\section{The Froeschl\'e Map} \label{The Froeschle Map}

\begin{figure*}[!hbt] \begin{center}
\includegraphics[height=14.cm,width=12.7cm]{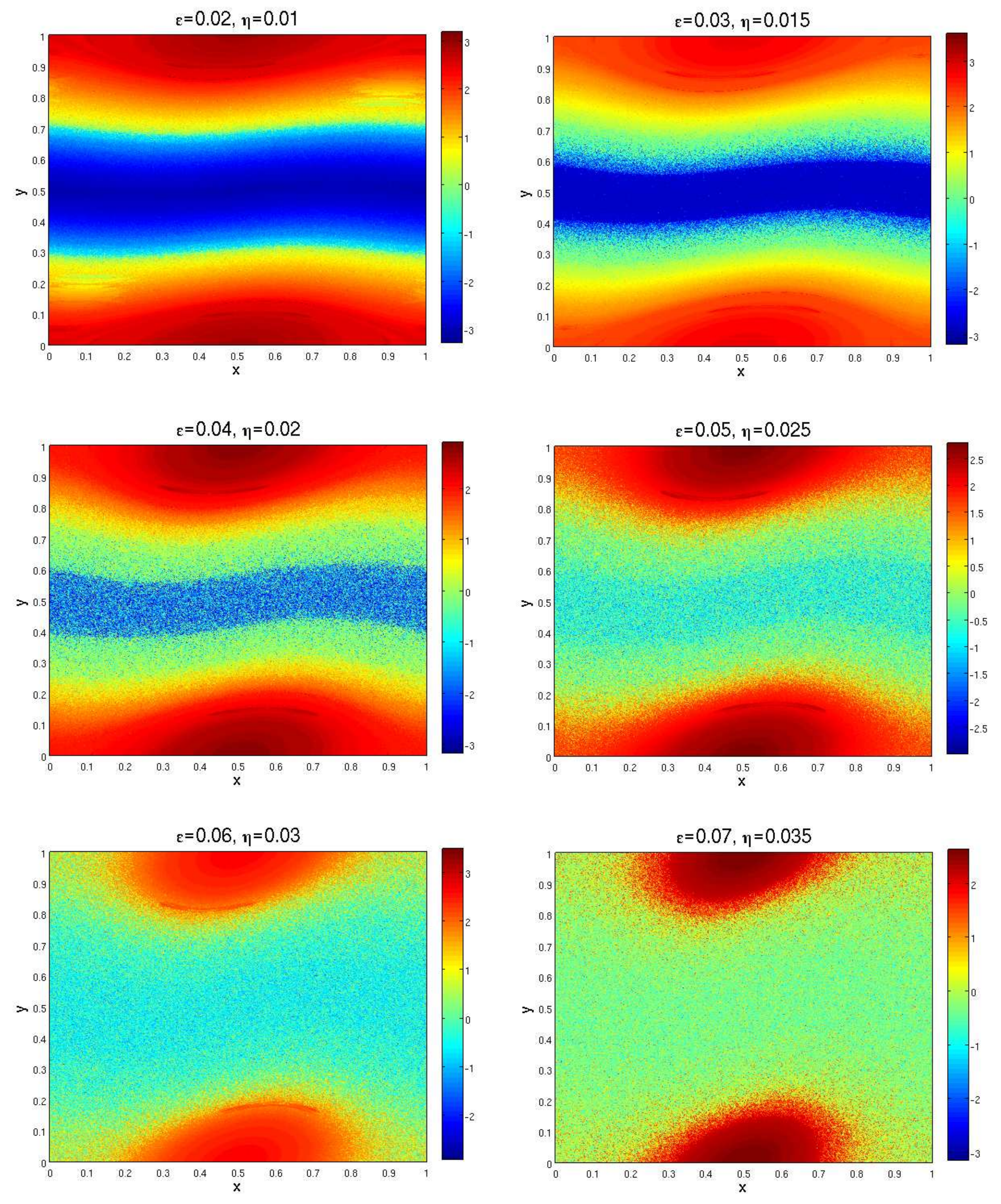}
\caption{Time averages of a single function $f_2 =  2 \cos (2 \pi y_1) + \cos (2 \pi y_1) \cos (2 \pi y_2) + \cos (12 \pi x_2) + \cos (12 \pi x_1)$ under the dynamics of the Froeschl\'e map Eq.\,(\ref{froeschlemap}), computed for a $500 \times 500$ grid of initial conditions taken on the phase space section $(x_2=0,y_2=0)$. The dynamics was run for $2 \times 10^5$ iterations. The values of $\e$ and $\eta$, linked by $\e=2\eta$, are indicated in each plot.}   \label{froeschle-eps=2eta-plots} 
\end{center} \end{figure*}

As our first higher dimensional example we consider the 4D, measure-preserving Froeschl\'e map \cite{froeschle} that consists of two standard maps
with a symplectic coupling: 
\begin{equation} \begin{array}{ll}
x_{1}' &= \; x_{1} + y_{1}  + \e_{1} \sin (2 \pi x_{1}) + \eta \sin (2 \pi x_{1} + 2 \pi x_{2})  \\
y_{1}' &= \; y_{1} + \e_{1} \sin (2 \pi x_{1}) + \eta  \sin (2 \pi x_{1} + 2 \pi x_{2})   \\
x_{2}' &= \; x_{2}  + y_{2} + \e_{2}  \sin (2 \pi x_{2}) + \eta  \sin (2 \pi x_{1} + 2 \pi x_{2})   \\
y_{2}' &= \; y_{2} + \e_{2} \sin (2 \pi x_{2}) + \eta  \sin (2 \pi x_{1} + 2 \pi x_{2}) 
\end{array} \label{froeschlemap} \end{equation}
where $(x_{1},y_{1},x_{2},y_{2}) \in [0,1]^{4}$. We set $\e_1=\e_2=\e$ and reduce our investigation to the case of two identical interacting standard maps. This map can be related to the standard map in two ways: for $\eta=0$ we have a system of two uncoupled standard maps, while for $\e=0$ one can introduce the new variables:
\[ \begin{array}{ll}
 u_1 = x_1 + x_2  \\
 v_1 = y_1 + y_2 \\
 u_2 = x_1 - x_2  \\
 v_2 = y_1 - y_2
\end{array} \]
which reduce the system to two maps: a standard map in coordinates $(u_{1},v_{1})$ with the parameter $2\eta$, and an integrable twist map in $(u_{2},v_{2})$. Given this structure, an interesting way of varying parameters of the map Eq.\,(\ref{froeschlemap}) is by letting  $\e=2\eta$. We slice the 4D phase space by a 2D section with fixed $(x_2,y_2)=(0,0)$, and consider a grid of $500 \times 500$ initial grid-points in $(x_1,y_1)$-space, plotting the time averages of a single function $f_2 =  2 \cos (2 \pi y_1) + \cos (2 \pi y_1) \cos (2 \pi y_2) + \cos (12 \pi x_2) + \cos (12 \pi x_1)$ obtained by running the dynamics for 200000 iterations. The results are shown in Fig.\,\ref{froeschle-eps=2eta-plots}. Note the departure from the known standard map's phase space due to increasing interaction between the maps. For small $\e$ values, the phase space  maintains its regular structure with only small and localized chaos similarly to standard map for small $\e$. The coloring here indicates two dimensional intersections of the real 4D \inv sets with the selected 2D phase space section. It appears that the global chaotic transition in the sense of vertical transport through the phase space occurs around $\e=2\eta=0.05$. 

\begin{figure}[!hbt] \begin{center}
\includegraphics[height=6.5cm,width=8.cm]{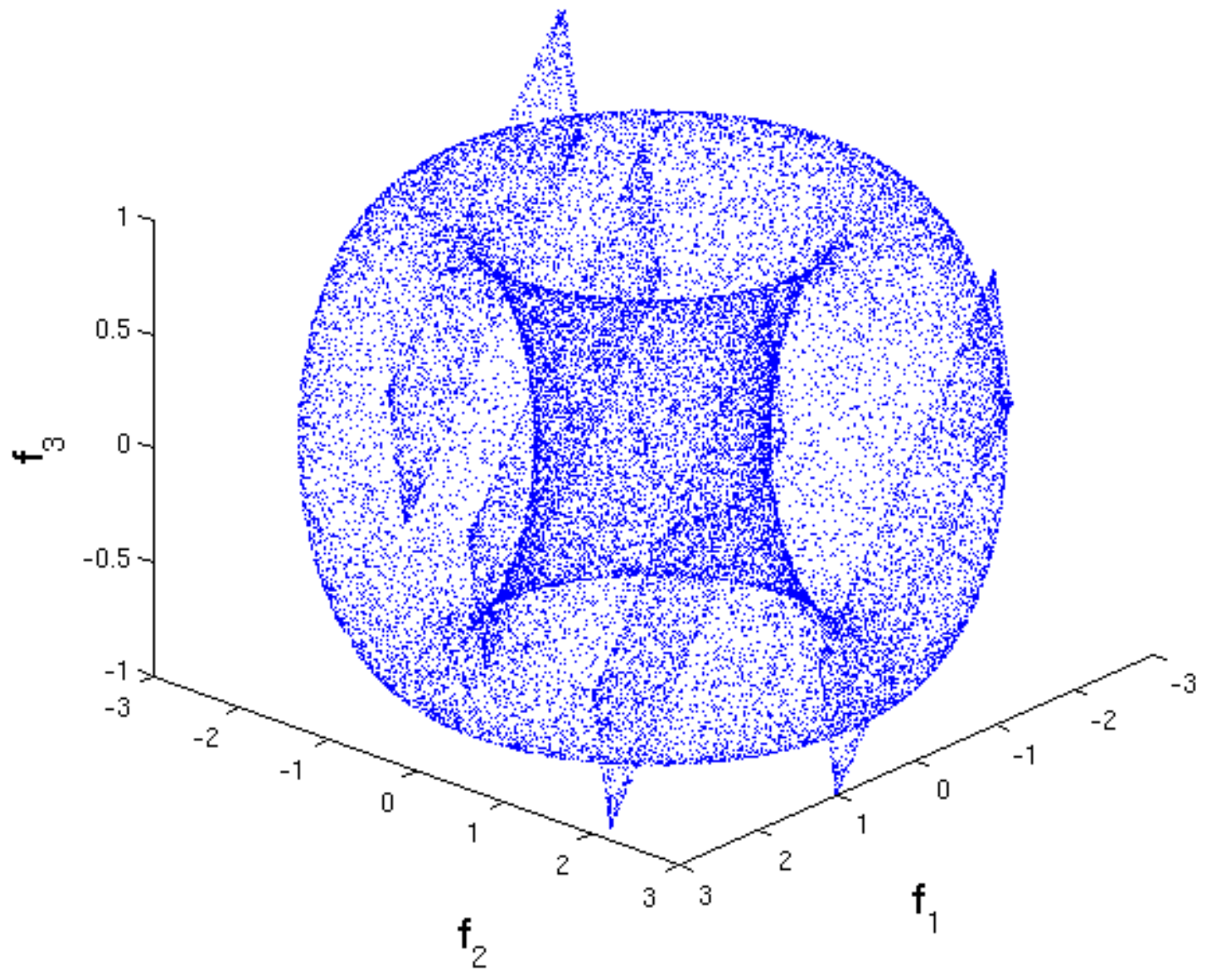}
\caption{Three-function MSP for the Froeschl\'e map Eq.\,(\ref{froeschlemap}) with $\e=\eta=0$. The grid of $13 \times  13 \times 13 \times 13$ initial points is constructed to visualize both resonances and irregular orbits 
   (see text). The dynamics is run for $10^5$ iterations. The used functions are: $f_1 =  2 \sin (2 \pi y_1) + \sin (2 \pi y_1) \cos (2 \pi y_2) + \cos (12 \pi x_1)$, 
   $f_2 =  2 \cos (2 \pi y_1) + \cos (2 \pi y_1)  \cos (2 \pi y_2) + \cos (12 \pi x_2) + \cos (12 \pi x_1)$ and $f_3 =  \sin (2 \pi y_2) + \cos (12 \pi x_2)$. } \label{froeschle-eps0eta0}  
\end{center}  \end{figure}

The structure of $Q_e$ for the Froeschl\'e map for $\e=\eta=0$ is clearly a product of two "rational combs" discussed earlier. This product is topologically a torus to which:
\begin{itemize} 
\item a line  of length $\frac{1}{q}$ is attached at every point  $(y_1,y_2)$ such that one of the $y_i$'s is irrational and the other rational, $\frac{p}{q}$ (with $\frac{p}{q}$ an irreducible fraction), 
\item a rectangle of sides $q_1,q_2$ is attached at every point $(y_1=\frac{p_1}{q_1},y_2=\frac{p_2}{q_2})$, where $\frac{p_1}{q_1}$ and $\frac{p_2}{q_2}$ are irreducible fractions.
\end{itemize}
Since two rectangles generically do not intersect in a $5$ dimensional Euclidean space, the appropriate embedding dimension for the Froeschl\'e map is five. To avoid self-intersections in the embedding, we would thus need to consider at least five independent functions. This information is important when choosing the number of functions that are used in clustering and approximation of the ergodic partition. However, since we can not graphically represent the 5D space results, we consider here three-function MSP visualized in 3D embedding. An example with $\e=\eta=0$ is shown in Fig.\,\ref{froeschle-eps0eta0}, obtained for a 4D grid $13 \times 13 \times 13 \times 13$ initial points as follows: 
\begin{itemize}
 \item the lattice-grid is set in full 4D phase space, and iterated using the Froeschl\'e map Eq.\,(\ref{froeschlemap}) with $\e=\eta=1$ for 10 iterations, in order to randomize the 
  points (starting from the uniform lattice-grid in the case of $\e=\eta=0$ might create trajectories that strongly overlap 
 \item from thus obtained  $13 \times 13 \times 13 \times 13 = 28561$ initial 4D points, set  $13 \times 13 \times 13 = 2197$ to have $(y_1=0,y_2=0)$, another $13 \times 13 \times 13$ 
to have $(y_1=\frac{1}{2},y_2=\frac{1}{2})$ and so on, depending on how many resonances are to be visualized 
 \item run the dynamics for $t_{\mbox{final}}=100000$ iteration starting from thus created ensemble of initial 4D points
 \item as the result, majority of points will account for quasiperiodic orbits, while the selected points will visualize the chosen resonances 
\end{itemize}
Note that in the Fig.\,\ref{froeschle-eps0eta0}, all the mentioned phase space features are visualized as expected. The chosen resonances for this case are  $(y_1=0,y_2=0)$,  $(y_1=\frac{1}{2},y_2=\frac{1}{2})$, $(y_1=\frac{1}{3},y_2=\frac{1}{3})$, and $(y_1=\frac{2}{3},y_2=\frac{2}{3})$.

\begin{figure*}[!hbt]  \begin{center} 
\includegraphics[height=12.cm,width=14.5cm]{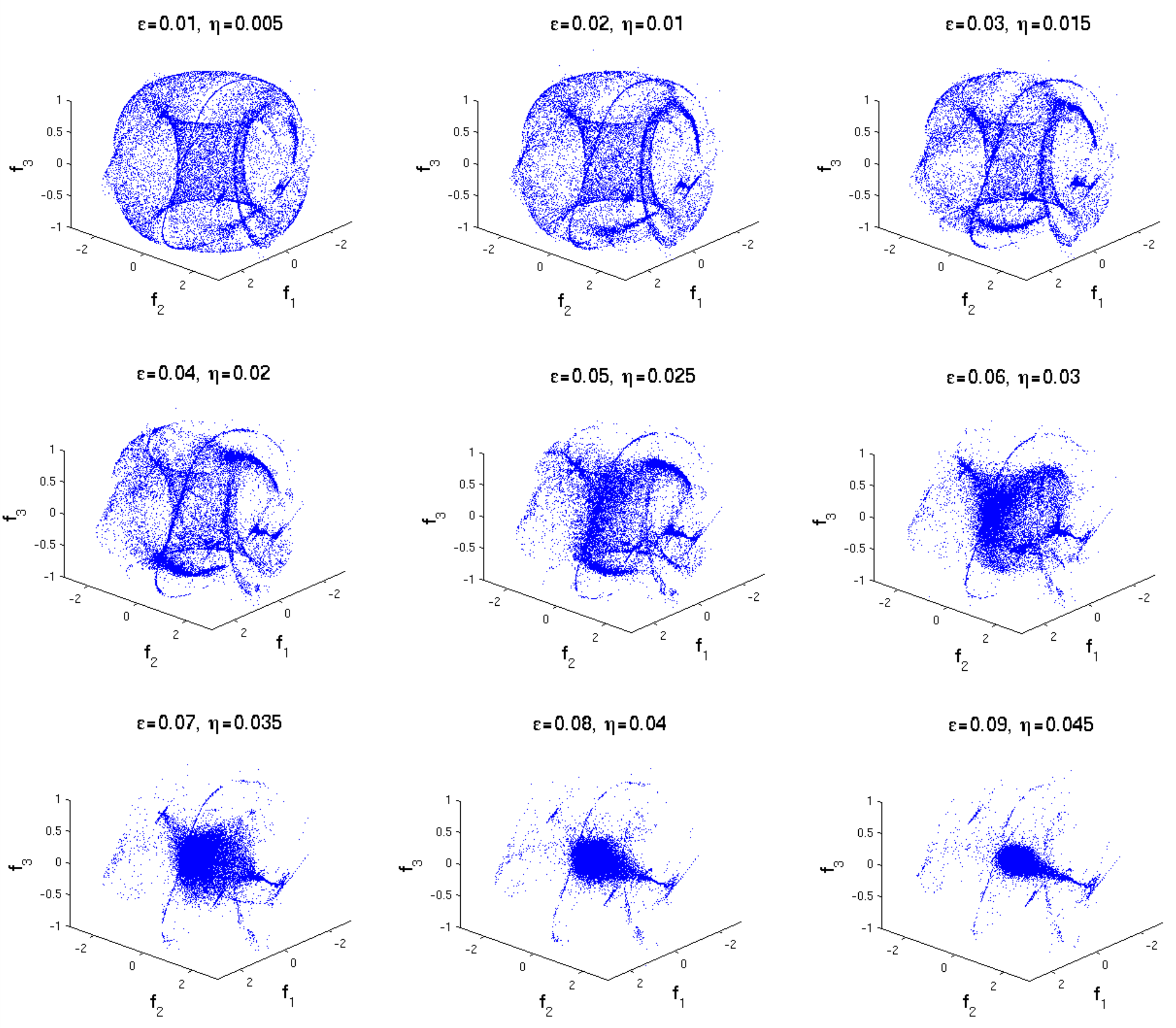}
\caption{Sequence of three-function MSP for the Froeschl\'e map Eq.\,(\ref{froeschlemap}). The grid of $12 \times 12 \times 12 \times 12$ initial points is constructed as for the Fig.\,\ref{froeschle-eps0eta0}. 
   The functions used are: $f_1 =  2 \sin (2 \pi y_1) + \sin (2 \pi y_1) \cos (2 \pi y_2) + \cos (12 \pi x_1)$, $f_2 =  2 \cos (2 \pi y_1) + \cos (2 \pi y_1)  \cos (2 \pi y_2) + \cos (12 \pi x_2) + \cos (12 \pi x_1)$ 
   and $f_3 =  \sin (2 \pi y_2) + \cos (12 \pi x_2)$, and the dynamics was run for $10^5$ iterations. The respective $\e=2\eta$ values are indicated in each plot.} \label{froeschle-eps=2eta-MSP} 
\end{center} \end{figure*}

We examine the three-function MSPs for the Froeschl\'e map with $\e=2\eta>0$, by employing 4D grid of $12 \times 12 \times 12 \times 12$ initial grid-points and the same number of total iterations. The ensemble of initial points with selected resonance points is constructed as above. We consider structural changes in the MSPs by changing the $\e$-value, as shown in Fig.\,\ref{froeschle-eps=2eta-MSP}. As the coupling intensity grows, the structure from the previous figure is destroyed, in a way similar to what observed for the standard map (cf. Fig.\,\ref{sequence3d}). In particular, note the different mechanisms of destruction of resonances and irrational orbits. The MSP's structure reports a given level of regularity in the phase space persisting for a certain range of coupling parameter strengths. This is a manifestation of KAM/resonance zone nature of this map, which (similarly to the standard map) maintains some \inv tori until the parameters exceed certain thresholds. From the MSP, we conclude that global merging of resonances occurs for value of $\e=2\eta$ between $0.05$ and $0.06$ (cf. phase space cross sections visualized in Fig.\,\ref{froeschle-eps=2eta-plots}). 

We conclude the Section by observing that MSP analysis as employed here allows investigations of systems much more complex than the standard map. As pointed out  previously, with an appropriate choice of functions one can reduce the complexity of a given system to geometrical features of the MSP, capturing the key dynamical details of the system in the form of MSP structure. Moreover, various changes in system's properties can be monitored this way (cf. Fig.\,\ref{froeschle-eps=2eta-MSP}) by observing the geometrical evolution of the MSP. A further application of this technique might be in the study of invariant sets of even higher dimensional dynamical systems, like the coupled maps on networks \cite{levnajictadic,mezic-dutoit} and  recently discovered maps that mimic quantum chaos \cite{horvattriangle}. \\[0.2cm]


\section{Extended standard map} \label{Extended Standard Map}

As a second higher dimensional example we consider the 3D extended standard map \cite{esm}, that represents a  generalization of the classical standard map. It is a volume-preserving action-action-angle map defined as:
\begin{equation} \begin{array}{lllc}
x' &= x + \e \sin (2\pi z) + \delta \sin (2\pi y)      \;\;\;  &[mod \; 1]  \\
y' &= y + \e \sin (2\pi z)                             \;\;\;  &[mod \; 1]  \\  
z' &= z + x + \e \sin (2\pi z) + \delta \sin (2\pi y)  \;\;\;  &[mod \; 1]
\end{array} \label{esm-eq} \end{equation}  
with the values in $[0,1]^3$. Its physical origin and the analytical properties are investigated in \cite{esm}. For $\delta = 0$ the map takes the form:
\begin{equation}  \begin{array}{lll} 
x' &= x + \e \sin (2\pi z)                     \;\;\;  &[mod \; 1]  \\
y' &= y + \e \sin (2\pi z)                     \;\;\;  &[mod \; 1]  \\
z' &= z + x + \e \sin (2\pi z)                 \;\;\;  &[mod \; 1]
\end{array} \label{esm-d0} \end{equation}
which keeps the planes $y-x=const.$ \inv under the dynamics, and reduces to the standard map in $x$ and $z$ coordinates (while $y$ behaves like another action coordinate). This  family of standard maps on diagonal planes is however broken for $\delta>0$ as the transport is allowed between the diagonal planes.  

\begin{figure*}[!hbt] \begin{center}
    \includegraphics[height=6.cm,width=14.cm]{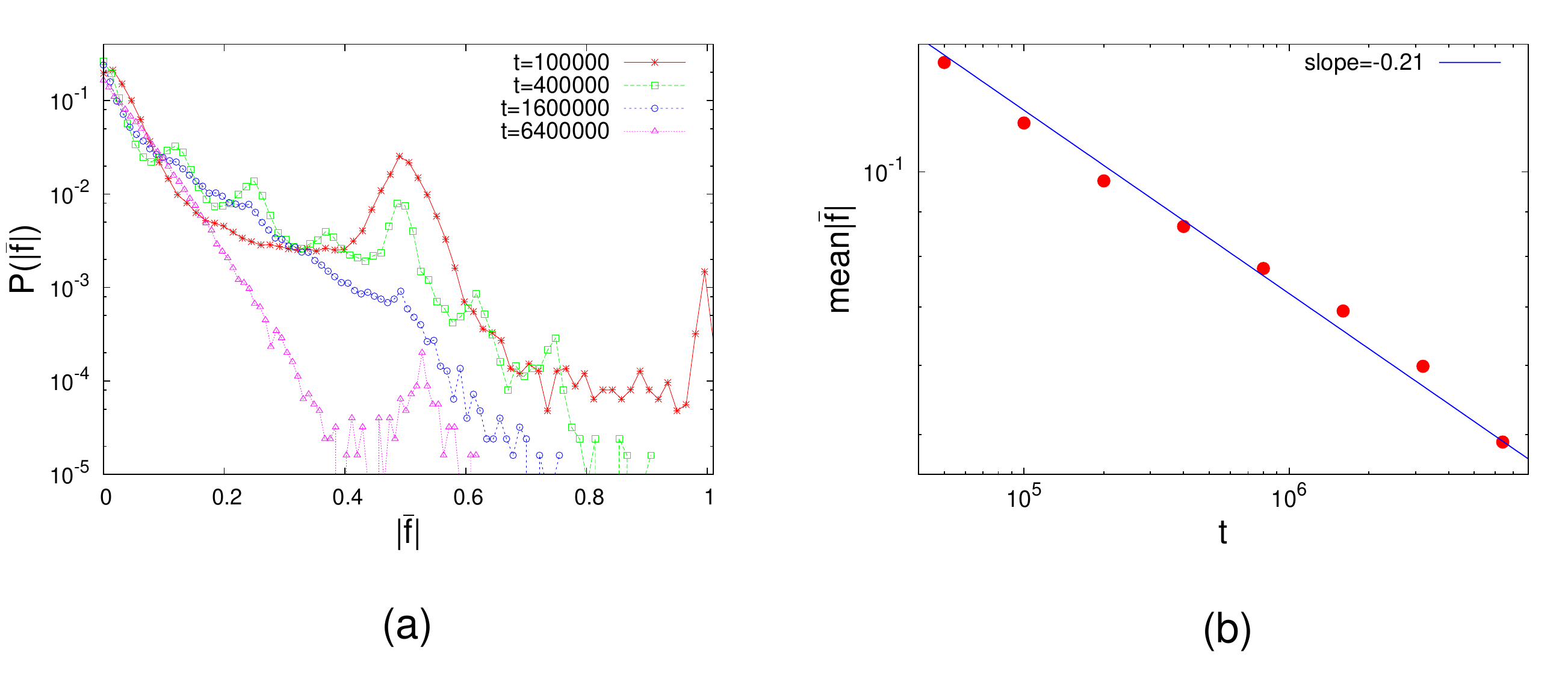}
\caption{Time average vectors $\mathbf{\bar f}^t=(f_1^t,f_2^t,f_3^t)$ are computed for extended standard map Eq.\,(\ref{esm-eq}) for the functions $f_1= \sin (2 \pi x)$, $f_2 = \cos(2 \pi y)$ and $f_3 = \sin(4 \pi y) \cos (4 \pi x) \cos(12 \pi z)$ with $\e=0.01$ and $\delta=0.001$ on the grid of $50 \times 50 \times 50$ random initial points for various numbers of iterations.  (a): time-evolution of the distribution of time average vector norms $|\mathbf{\bar f}^t|$; (b): the distribution's mean value as function of time $t$, fitted with the slope of -0.21.} \label{esm} 
\end{center} \end{figure*}

It was conjectured in \cite{esm} that the extended standard map Eq.\,(\ref{esm-eq}) is ergodic for small positive values of perturbations $\e$ and $\delta$. We bring additional evidence to this claim in Fig.\,\ref{esm}, where we examine the way time average vectors shrink to zero with time-evolution of this map for $\e=0.01$ and $\delta=0.001$ (equivalently to what was done in Fig.\,\ref{clustershrinking}). Time average vectors $\mathbf{\bar f}^t=(f_1^t,f_2^t,f_3^t)$ are computed for extended standard map Eq.\,(\ref{esm-eq}) with $\e=0.01$ and $\delta=0.001$. The grid of $50 \times 50 \times 50$ initial points was used, with randomized points as in the previous Section. The functions $f_1= \sin (2 \pi x)$,$f_2 = \cos(2 \pi y)$ and $f_3 = \sin(4 \pi y) \cos (4 \pi x) \cos(12 \pi z)$ are considered. In Fig.\,\ref{esm}a we show the distributions of time average vector norms $|\mathbf{\bar f}^t|$ defined as in Eq.\,\ref{tav-norm}, in function of time $t$ (cf. Fig.\,\ref{clustershrinking}a). It is interesting that at $t=400000$, the structure of resonances is present due to a quasi-two-dimensional nature of the map causes a multi-modal distribution of time averages. This multi-modality disappears with the higher number of iterates and the distribution assumes an exponential shape, with variance and mean tending to zero. In Fig.\,\ref{esm}b the mean values of distributions are shown in function of number of iterations $t$, and fitted with the slope of -0.21. This suggests an average convergence rate of $t^{-0.21}$, slower than for the case of strong chaos characterized by $t^{-0.5}$ (cf. Fig.\,\ref{clustershrinking}b).

This seems to confirm the mentioned ergodic hypothesis from the computational prospective. It also agrees with the result proved in \cite{esm} stating that no invariant two-dimensional surface persists in this map for any positive perturbation value, implying that the map allows a global transport throughout the phase space at a small non-zero perturbation. This consideration demonstrates our method to be useful in the context of numerical investigations related to ergodic properties of dynamical systems as the one discussed above.


\section{Conclusions} \label{Conclusions}

We presented the computational realization and theoretical extension of an \inv set visualization method based on ergodic partition theory suggested in \cite{mezicwiggy}. We defined the ergodic quotient space obtained by associating an ergodic set with a point. Embeddings of the ergodic quotient space into Euclidean space, called Mesochronic Scatter Plots (MSP's) were realized using time averages of observables on the phase space. The time averages  were computed and visualized using a coloring scheme that we named a Mesochronic Plot for a variety of measure-preserving maps. The time average convergence issues have been considered.  A simple algorithm for approximation of the ergodic partition was developed from multi-functional MSP's by dividing the time average vectors space into cubical cells. Approximation of ergodic partition  structure in the phase space was shown for various numbers of functions. By studying the standard map with known properties, we were able to confirm the visualization results within the limits of numerical precision. The extent of the method's applicability was illustrated on 4D Froeschl\'e map and 3D extended standard map, giving new insights into dynamical structure of these systems. Ergodic invariant sets in higher dimensional systems can be visualized using our method by obtaining their intersections with two-dimensional surfaces of choice.

In the paper to follow \cite{levnajicmezic2} we show how periodic sets and resonances can be graphically visualized according to their periodicity and the phase space structure, by the use of \textit{harmonic time averages} that extend the concept of time average described here. Both the method presented here and the method in the follow-up paper are related to eigenspace structure of the Koopman operator.

The further improvement of the method can be obtained by optimization of clustering techniques beyond the simple cell division exposed here \cite{budisic}. The selection of optimal functions for embedding is a wide-open question. Moreover, a more detailed geometric analysis of ergodic quotient space might yield additional insights into the dynamics, and indicate a way to construct better algorithms for clustering of time average vectors and ergodic partition approximation. The integrable maps studied here have interesting mathematical structure that is non-smooth but still in some sense regular. For example, the unperturbed standard map has ergodic quotient space structure of a unit circle with an interval of length $\frac{1}{q}$ attached at every rational point $\frac{p}{q}$.  

The method can also be applied to the  continuous-time systems for which the ergodic theory results are equally valid, and hence the results shown here apply directly. Of course, the computation of time averages for a continuous-time system is far more numerically demanding.\\[.3cm]


\n {\bf Acknowledgments.} This work was supported the DFG through the project FOR868, by the AFOSR grant numbers F49620-03-1-0096 and FA9550-09-1-0141, and by the national Program P1-0044 (Slovenia). Thanks to prof.s B. Tadi\'c, T. Prosen and A. Pikovsky for useful comments. Thanks to U. Vaidya and G. Cristadoro for constructive discussions. Special thanks to R. Krivec at J. Stefan Institute for maintenance of the computing resources where most of the numerical work was done. Part of this work was done during ZL's stay at Univ. California Santa Barbara, and part during his stay at Dept. of Theor. Physics, J. Stefan Institute, Ljubljana, Slovenia.

\begin{scriptsize}
  \end{scriptsize}

\end{document}